\documentclass[%
 aip,
jcp,
 amsmath,amssymb,
 reprint,%
]{revtex4-1}

\usepackage{graphicx}
\usepackage{dcolumn}
\usepackage{bm}
\usepackage{longtable}
\usepackage{hyperref}

\usepackage[utf8]{inputenc}
\usepackage[T1]{fontenc}
\usepackage{mathptmx} 

\begin{document}

\title{First-principles Equations of State and Structures of Liquid Metals in Multi-megabar Conditions}

\author{Shuai Zhang} 
 \email{szha@lle.rochester.edu}
\affiliation{Lawrence Livermore National Laboratory, Livermore, California 94550, USA}
\affiliation{Present address: Laboratory for Laser Energetics, University of Rochester, Rochester, New York 14623, USA}%
\author{Miguel A. Morales}%
 \email{moralessilva2@llnl.gov}
\affiliation{Lawrence Livermore National Laboratory, Livermore, California 94550, USA}


\begin{abstract}
Liquid metals at extreme pressures and temperatures
are widely interested in the high-pressure community.
Based on density functional theory molecular dynamics,
we conduct first-principles investigations on
the equation of state (EOS) and structures of four metals (Cu, Fe, Pb, and Sn)
at 1.5--5 megabar conditions and 5$\times10^3$--4$\times10^4$ K.
Our first-principles EOS data enable evaluating the performance of four EOS models in predicting Hugoniot
densities and temperatures of the four systems.
We find the melting temperature of Cu is 1000--2000 K higher
and shows a similar Clapeyron slope, in comparison to those of Fe.
Our structure, coordination number, and diffusivity analysis indicates all the four liquid metals form similar simple close-packed structures.
Our results set theoretical benchmarks for EOS development
and structures of metals in their liquid states and under
dynamic compression. 
\end{abstract}

\maketitle

\text{LLNL-PROC-783211}

\section{\label{sec:intr}Introduction}

Metallic liquids widely exist in nature
at planetary interiors, such as the outer part of
the Earth's core which is made of iron-rich fluids
under multi-megabar (Mbar) pressures and thousands-Kelvin
temperatures. Essentially, convection of the conductive
fluids has been generating a magnetic 
field that is crucial to the planet's habitability.
Accurate understandings of the structure,
equation of state, and transport properties of
the high-pressure liquids are crucial to geophysical
modeling of the dynamics and chemistry
and deciphering the energy budget, geomagnetism,
and formation and evolution of the planet.~\cite{Olson2007}

Liquid forms of matter at extreme conditions are also ubiquitously generated in laboratory during shock experiments. Typically, the temperature-pressure conditions probed 
in the experiments are along the Rankine-Hugoniot 
curve~\cite{Meyers1994book}, which is much steeper 
than the melting curve
and therefore can easily produce high-pressure liquids. 

However, as a result of interplay between complexities in
atomistic and electronic structures, the study of high-pressure 
liquids has posed grand challenges to both theory and experiment~\cite{McMillan2007}.
Over the past few decades, experimental studies 
have been relying on x-ray/neutron scattering and
optical probes combined with 
static compression techniques,
such as those using a high-pressure vessel~\cite{Inui2007},
a multi-anvil press~\cite{Tsuji1989,Yamada2011},
or a diamond-anvil cell (DAC)~\cite{Shen2004,Eggert2002},
to measure the structure and properties of liquids under pressure.
These experiments, usually performed at up to a few tens gigapascal (GPa) in pressure and
a few thousands Kelvin in temperature, 
have revealed many
interesting physics of 
structural and electronic phase transitions
in a wide variety of materials.
For example, evidences were found 
on the existence of liquid-liquid phase (LLP) transitions 
in single-element substances of phosphorus~\cite{Katayama2000},
silicon~\cite{Sastry2003,McMillan2005,Beye2010,Vasisht2013},
germanium~\cite{Bhat2007}, 
 gallium~\cite{Tien2006},
nitrogen~\cite{Jiang2018},
and hydrogen~\cite{Dzyabura2013,McWilliams2016,Ohta2015,Zaghoo2016}
as well as in oxides, molecules, and alloys~\cite{Brazhkin2003,Kobayashi2016},
and on metallization in liquid selenium, sulfur, iodine, nitrogen, and hydrogen~\cite{Weir1996,Brazhkin2006,Jiang2018}.

In recent years, techniques that combine ultra-bright, 
high-energy x-ray with dynamic compression 
have been developed in places such as
the Matter at Extreme Conditions end station of the 
Linac Coherent Light Source x-ray free-electron laser~\cite{Gorman2018}
and the Dynamic Compression Sector at Advanced Photon Source~\cite{Wang2019}.
These advances have enabled {\it in situ} structural determination of liquids in laboratory 
along with equations-of-state measurements for 
matter at extreme conditions.

Theoretically, a natural way of simulating materials 
at finite temperatures is by doing molecular dynamics (MD).
In these simulations, atoms are treated as particles 
that interact with potentials in certain forms.
The potentials are usually empirical and chosen either in 
analytic or numerical forms 
(so called ``classical MD'').
Classical MD approaches
have shown great usefulness
in simulating materials 
in a temporal scale (e.g., nanoseconds) 
that is close to that experienced by materials in
shock compression 
experiments or during meteorite impacts~\cite{Melosh2007}.
The considerable spacial scale (e.g., millions 
of atoms) of classical MD simulations provides
rich information about the response of materials
at high pressure and temperatures.~\cite{Bolesta2017,CarvajalJara2009}
However, due to the temperature and pressure
dependence of electronic interaction,
accuracy and transferability of the empirical 
potentials and reliability of the classical MD results
are typically questionable.~\cite{Glosli1999,Wu2002}
And one has to rely on first-principles MD~\cite{Ganesh2009,Boates2009,Morales2010,Pierleoni2016}, such as those based on
density functional theory (DFT)~\cite{ks1965,hk1964,Mermin1965}, 
in order to acquire
a more precise description of the electronic
interaction and properties of materials.

As the state-of-art quantum mechanical approach
for condensed matter studies,
DFT simplifies the many-body interaction in real materials
into a single-particle mean field problem, with an
effective potential including an exchange-correlation
term that is subject to users' choices in their calculation.
This has enabled accurate simulations to the electronic
level that is computationally feasible.
DFT-MD has been widely used to set theoretical
benchmarks for the equation of state (EOS), 
structures, and properties
of a wide variety of materials at cold to warm dense conditions~\cite{Morales2013,Militzer2018,Zhang2016,Zhang2016b,Zhang2017,Zhang2017b,Zhang2018,Zhang2018b,Zhang2019,Li2018,Millot2020},
in particular liquids in multi-Mbar 
conditions. These calculations, in synergy
with the ongoing experimental developments,
are expected to unveil the extreme physics of
the matter and provide important inputs
for developing better potentials for large-scale MD simulations and building reliable EOS models for hydrodynamic simulations.

In this paper, we report DFT-MD simulations of four
metals: iron (Fe), copper (Cu), lead (Pb), and tin (Sn).
The calculations are at 1.5--5 Mbar 
around the respective Hugoniot curve of the materials.
The discussions are focused on the EOS, shock Hugoniot,
and the atomistic structure. We also briefly talk about
transport properties in connection to the structural results.
Our results indicate that all the four systems, when 
in their liquid form and at the conditions considered in
this work, show close-packed simple structures.

\section{\label{sec:method}Computational details}

\begin{table*}
\caption{\label{tab:table1} Parameters chosen in our DFT-MD simulations.}
\begin{ruledtabular}
\begin{tabular}{ccccc}
& Fe & Cu & Pb & Sn \\
\hline 
atoms/simulation cell & 128 & 128 & 256 & 256 \\
$E_\text{cutoff}$ (eV) & 800 & 800 & 500 & 400 \\
ENMAX (eV) & 267.882 & 295.446 & 97.973 & 103.236 \\
$r_\text{c}$ (Bohr) & 2.3 & 2.3 & 3.1 & 3.0 \\
valence electrons & 8 & 11 & 4 & 4 \\
Pseudopotential & PAW, 22Jun2005 & PAW, 06Sep2000 & PAW, 08Apr2002 & PAW, 08Apr2002 \\
Exchange-correlation functional & \multicolumn{4}{c}{PBE} \\
$k$-point & \multicolumn{4}{c}{(1/4, 1/4, 1/4)2$\pi/a$ ($a$ being the lattice constant of the cubic simulation cell)} \\
\end{tabular}
\end{ruledtabular}
\end{table*}

All our MD calculations are conducted within DFT and using the Vienna Ab-initio Simulation Package
({\footnotesize VASP})~\cite{kresse96b}.
The setup of the simulations are summarized in Table~\ref{tab:table1}.
For simplicity, we implement the ``mean-value'' $k$ point~\cite{Baldereschi1973} of $(1/4,1/4,1/4){2\pi}/{a}$, where $a$ denotes the size of the cubic simulation cell, to sample the entire Brillouin zone. 
We use the Perdew-Burke-Ernzerhof (PBE) exchange-correlation functional, a projected augmented wave (PAW)~\cite{Blochl1994} pseudopotential, and time steps of 1.5--5.2 fs that depend on the temperature and the density.

For Fe and Cu, we start each simulation from a 
128-atoms cell constructed by $4\times4\times4$ times 
the 2-atom body-centered-cubic (bcc) unit cell.
The simulations of Sn and Pb use cubic cells each 
containing 256 atoms and starting with representative
liquid snapshots from simplified ``warm-up'' calculations.
The setup for the ``warm-up'' calculations is similar to the main, productive ones but the number of electronic bands are smaller.
This makes the calculation faster while still generating  
reasonable structural snapshots in the number of a few thousands that consist the MD trajectory.
The starting configuration for the ``warm-up'' calculations are $4\times4\times4$ times 
the 4-atom face-centered-cubic (fcc) unit cell.

We use a Nos\'e thermostat~\cite{Nose1984} to generate MD trajectories in canonical ($NVT$, i.e., constant umber of atoms, constant volume, and constant temperature) ensembles.
Each MD trajectory consists 12000--20000 steps for Sn and Pb, and 1000--8000 steps for Cu and Fe. When calculating the EOS and analyzing the structures, we disregard the beginning 20\% and perform block averaging over the remaining part of each MD trajectory, in order to make sure the analysis are for the system in equilibrium. The ion kinetic contributions to the pressure and energy are included in the EOS following an ideal gas model.
We perform convergence tests on the cell size, basis set
cutoff, and $k$ sampling point/grid, and found the EOS, structure, and diffusivity results to be the same within the standard errorbar of our data.



\begin{table}
\caption{\label{tab:table2} Initial conditions determined by DFT calculations and used for constructing Hugoniots in this work.}
\begin{ruledtabular}
\begin{tabular}{ccccccc}
& Fe & Cu & Cu & Pb & Pb & Sn \\
& bcc & fcc & bcc & bcc & fcc & $\beta$ \\
\hline 
$\rho_0$ (g/cm$^3$) & 7.877 & 8.96 & 8.96 & 11.34 & 11.34 & 7.287 \\
$E_0$ (eV/atom) & -8.228 & -3.726  & -3.688 & -3.517 & -3.558 & -3.785  \\
$P_0$ (GPa) & -6.60 & 3.73 & 0.52 & 1.91 & 2.15 & 2.58 \\
\end{tabular}
\end{ruledtabular}
\end{table}

Using the EOS data from the DFT-MD calculations, we determine the pressure-density-temperature shock Hugoniot curves. This is done via the Rankine-Hugoniot equation~\cite{Meyers1994book} $\mathcal{H}=E-E_0-(P+P_0)(V_0-V)/2=0$, where ($E, P, V$) and ($E_0, P_0, V_0$) denote the total internal energy, pressure, and volume of a sample under steady shock and in the initial state, respectively. 
The Hugoniots that are obtained by first fitting the EOS along 
each isotherm (isochore) using cubic splines and then determining the pressure, 
energy, and volume (temperature) conditions at which the Hugoniot 
equation is satisfied.
The values for the initial energy and pressure used in this work are summarized in Table~\ref{tab:table2}.
We obtain these numbers by performing DFT calculations using the ambient structure
of each metal, i.e., fcc at 8.96 g/cm$^3$ for Cu, ferromagnetic bcc at 7.877 g/cm$^3$ for Fe, bcc at 11.34 g/cm$^3$ for Pb, and $\beta$-tin at 7.287 g/cm$^3$ for Sn.
An additional bcc (fcc) structure with the same density for Cu (Pb) was tested to check the dependence of the Hugoniot on the initial condition.

We calculate the radial distribution function $g(r)$ by analyzing inter-atomic distances along MD trajectories. The structure factor is obtained from $g(r)$ according to the definition 
$S(k)=1+4\pi n\int_0^{\infty}r^2\frac{\sin(kr)}{kr}[g(r)-1]dr$, where $n$ is the density in units of atoms/volume. We choose the number of bins to be $\sim$100 when calculating $g(r)$ and a bin size of 0.005 \AA$^{-1}$ when calculating $S(k)$. We have tried other bin sizes and found the $g(r)$ and $S(k)$ profiles are not sensitive to those variations. We also calculate $S(k)$ from the DFT-MD trajectories according to its definition $\left<\rho^*(k)\rho(-k)\right>$, where $\rho(k)=\int\exp(i{\bf k}\cdot{\bf r})d{\bf r}$ is the electron density,
and found $S(k)$ results obtained in the two separate ways are consistent with each other.


\section{\label{sec:results}Results and Discussion}
\begin{figure*}
\includegraphics[width=0.4\textwidth]{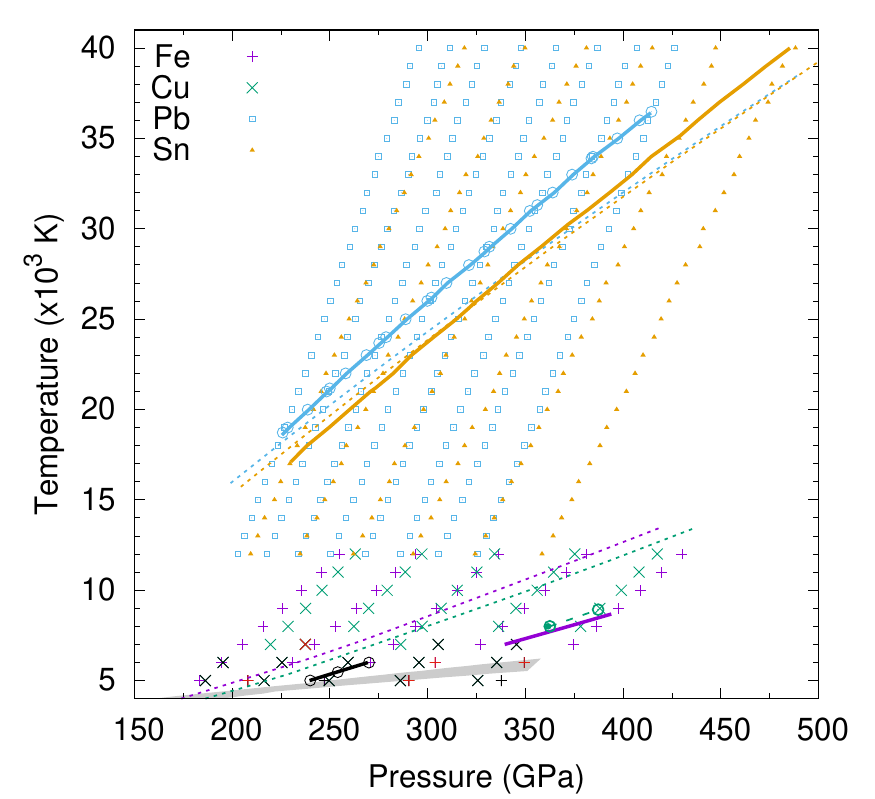}
\includegraphics[width=0.4\textwidth]{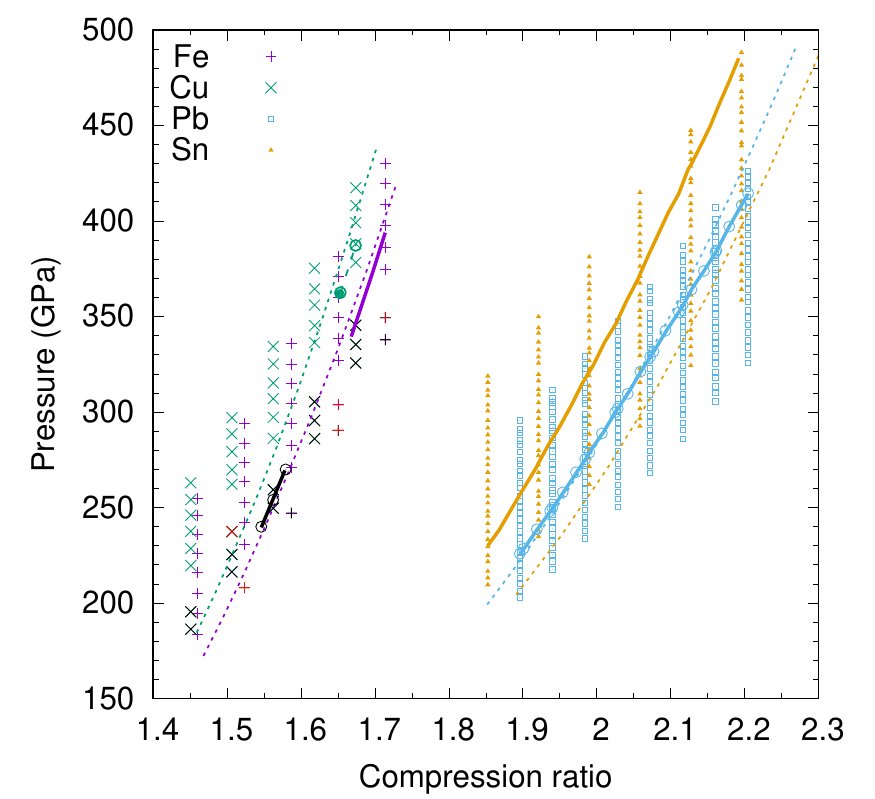}
\caption{\label{fig:eos} (left) Temperature-pressure and (right) pressure-compression ratio representation of the equations of state in this study. Symbols denote the EOS data points. Solid and dashed curves are the principle Hugoniot from our first-principles calculations and LEOS models, respectively. Black- and red-colored symbols denote simulations that show solidification and instability features respectively, according to the analysis of the mean-square displacement. The compression ratio is with respect to the ambient density of the corresponding metal, i.e., 7.877 g/cm$^3$ for bcc iron, 8.96 g/cm$^3$ for fcc copper, 11.34 g/cm$^3$ for bcc lead, and 7.287 g/cm$^3$ for $\beta$ tin. The open circles along the Cu (Pb) Hugoniots are Hugoniot conditions corresponding to a bcc (fcc) initial structure with the same density as that used for the fcc initial structure. The gray shaded area in the left panel denotes the recently measured melting curve of iron in resistance-heated DAC~\cite{Sinmyo2019}. The shock-wave data for iron melting temperature scatter above the DAC curve by $\sim$500--1700 K~\cite{Brown1986,Yoo1993,Nguyen2004}.}
\end{figure*}

\begin{figure*}
\includegraphics[width=0.4\textwidth]{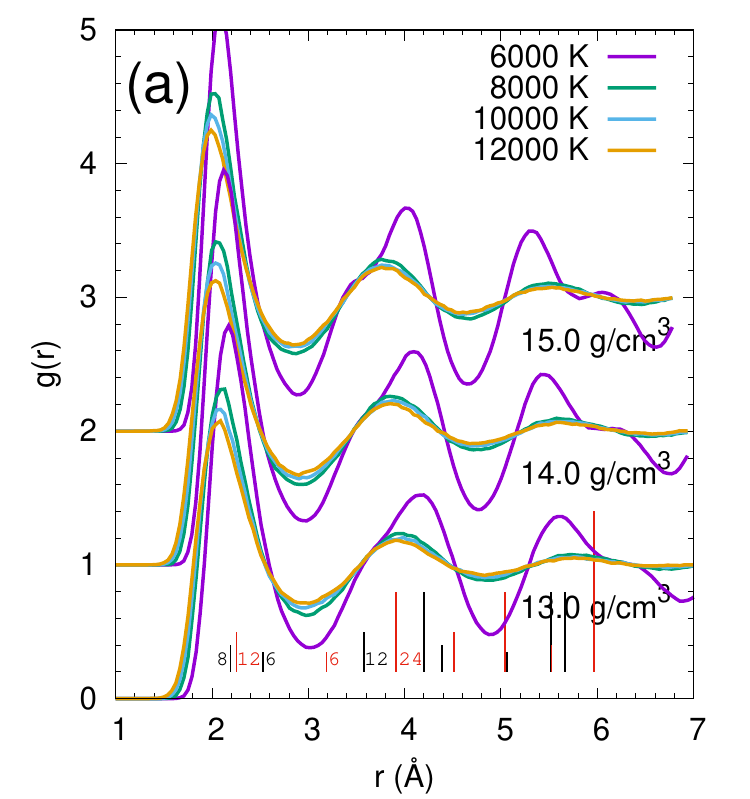}
\includegraphics[width=0.4\textwidth]{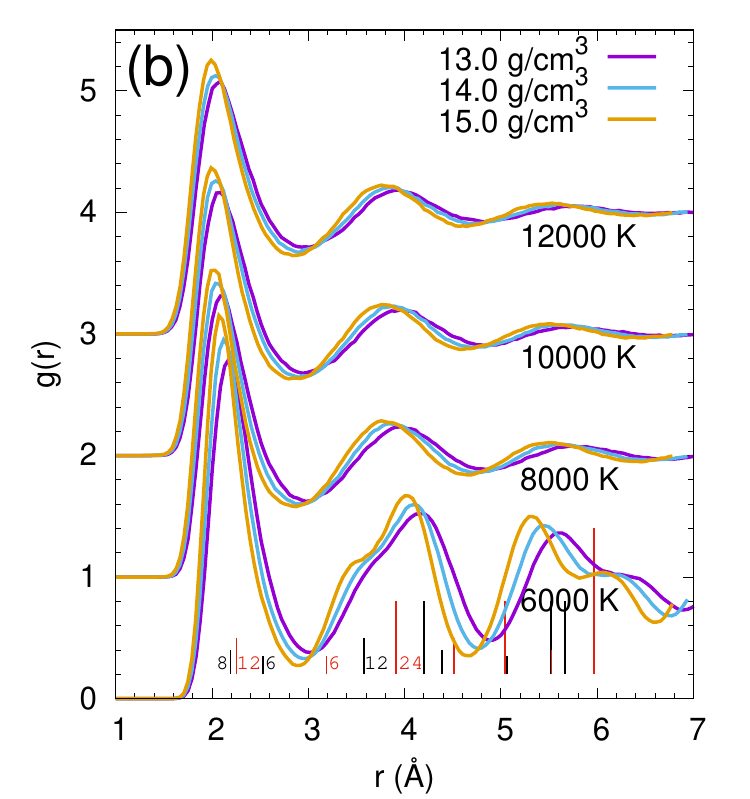}
\caption{\label{fig:strucCu} Pair-correlation function of copper at various temperatures and densities. Different isochores (a) and isotherms (b) have been shifted apart for clarity. Positions of the up-to-eighth (seventh)-nearest-neighbor shells of a perfect body (face)-centered-cubic crystal at 13.0 g/cm$^3$ are also shown with black (red) vertical bars for comparison. The height of bars are scaled by the coordination number of the corresponding shells, which are labeled for the first few shells.}
\end{figure*}

\begin{figure*}
\includegraphics[width=0.4\textwidth]{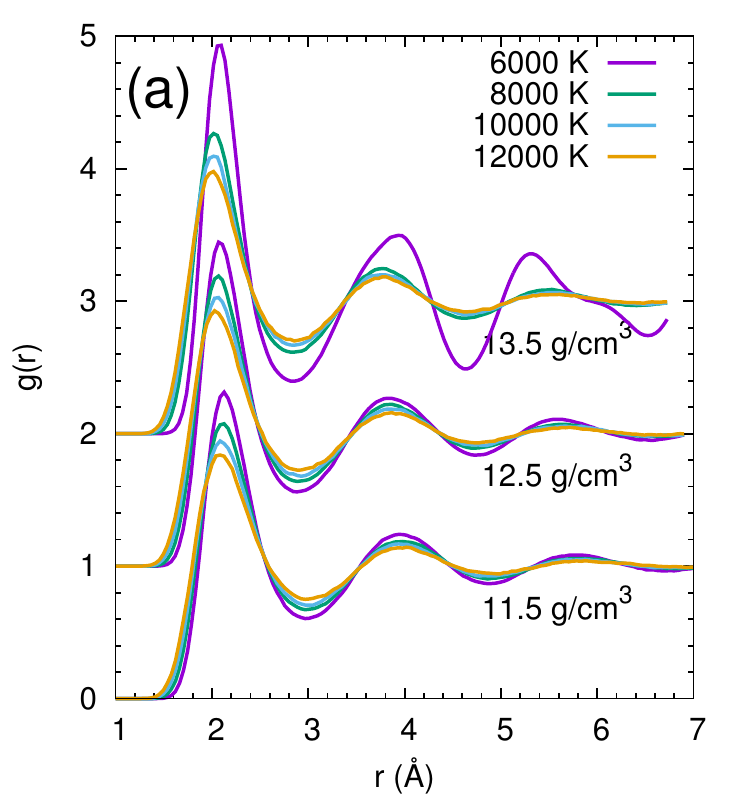}
\includegraphics[width=0.4\textwidth]{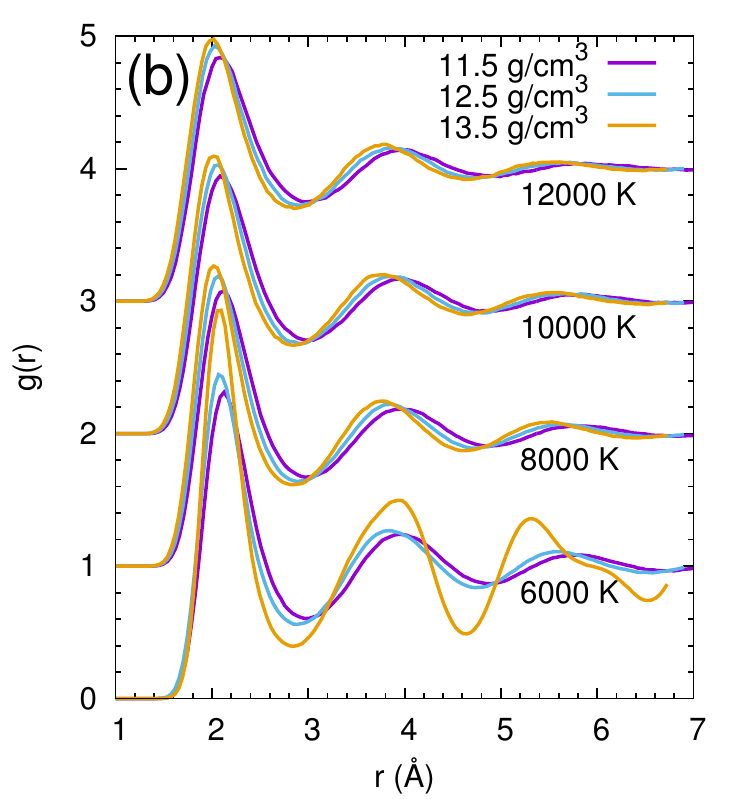}
\caption{\label{fig:strucFe} Pair-correlation function of iron at varies temperatures and densities. Different isochores (a) and isotherms (b) have been shifted apart for clarity.}
\end{figure*}

\subsection{\label{subsec:eos}EOS and Hugoniot}

Our first-principles EOS and Hugoniot for the four metals are summarized in Fig.~\ref{fig:eos}.
We also show Hugonoits predicted by LEOS models (LEOS 260 for Fe, LEOS 290 for Cu, LEOS 820 for Pb, and LEOS 500 for Sn) to evaluate their performances
by comparing to our DFT-MD results.

Our EOS results show that, around 1.5--5.0 Mbar, the Hugoniot temperatures of Pb and Sn is higher than that of Fe and Cu. This is explained by the larger compression ratio of Pb and Sn than Fe and Cu, which has to be compensated by a larger increase in the internal energy according to the Hugoniot equation $E-E_0=(P+P_0)(V_0-V)/2$. The larger compression ratio of Pb and Sn is consistent with their smaller bulk moduli~\footnote{https://periodictable.com/Properties/A/BulkModulus.v.html} and larger thermal expansion coefficients~\footnote{https://periodictable.com/Properties/A/ThermalExpansion.html} than those of Fe and Cu.

Along the isochores, our DFT-MD data show discontinuity in pressure
at densities between 12.0--13.5 g/cm$^3$ for Fe and 
13.0--15.0 g/cm$^3$ for Cu.
This is a signature that the simulation temperature is near the melting
curve. We use the mean-square displacement (MSD) 
as a criteria for judging whether the system remains
an equilibrium fluid, a crystalline solid (shown with 
black symbols in Fig.~\ref{fig:eos}),
or some meta-stable (red symbols in Fig.~\ref{fig:eos}) state.
This analysis gives melting temperatures of Fe that are
consistent with recent experiments~\cite{Sinmyo2019}.
Our results also show the melting temperature of Cu is
1000--2000 K higher than that of Fe in the pressure range
of 2--3.75 Mbar. In comparison to the comparison 
by Japel {\it et al}.~\cite{Japel2005prl} at up to 1 Mbar,
our results indicate the Clapeyron slopes of the melting curves of Fe and Cu
become more similar in the regime of 1.75--3.75 Mbar
than below 1 Mbar. 

In comparison to predictions by the LEOS models, our results indicate that 
LEOS 290 (260) predicts Cu (Fe) to be slightly harder
along the Hugoniot at pressures above 2 Mbar
and more so for Cu in the solid regime.
On the contrary, LEOS 500 predicts Sn 
to be softer than our first-principles predictions
and LEOS 820 predictions for the pressure-density Hugoniots of Pb 
are very similar to our DFT-MD results.
For the temperature-pressure Hugoniot, the LEOS 500 predictions for Sn are similar to our DFT-MD data, while LEOS 260 (290) predictions for Fe (Cu) are higher and LEOS 820 predictions for Pb are lower by up to 4000 K ($\sim$40\% for Fe, $\sim$30\% for Cu, $\sim$10\% for Pb) in the pressure range of 2--4 Mbar.

The LEOS models are constrained by experimental Hugoniot data (up to 
1.5-3.7 Mbar for Fe, Cu, Pb, and Sn)~\cite{MarshLASL1980} and subject to errors due to the
choice in the value of thermodynamic properties (e..g, Debye temperature and
Gr\"uneissen parameter) in the models for ion thermal  and electron thermal (e.g., Thomas-Fermi or Purgatorio) contributions to the free energy,
in particular at pressures where no experiment is available.
This could be a major reason for the differences between the Hugoniots predicted by the LEOS models and our calculations.
We note that the use of exchange-correlation functional in DFT to approximate the electronic interaction in real materials can also contribute to the differences, because of the non-uniformity of electrons density distribution due to the difference between the more localized inner-shell orbitals and the more extended valence and conduction states.
Other factors, such as pseudopotentials and electron relativistic effects may also be critical when considering heavy elements~\cite{Ahuja2011} and high pressures.
However, previous work have indicated DFT can predict high-pressure EOS that agree remarkably well with experiments. Examples include Cu under ramp compression to terapascal conditions~\cite{Fratanduono2020} and the ground-state isotherm of a wide variety of materials up to 1 Mbar~\cite{Soderlind2018}.
The thousands-Kelvin temperatures being studied are also much higher than the typical Curie temperature of metals, therefore the electron correlation effects could be less a problem than at the ground state.
However, detailed discussion on this is beyond the scope of this work and will be addressed, for the case of Sn, in a separate publication~\footnote{S. Zhang and M. Morales, in preparation.}.

\begin{figure*}
\includegraphics[width=0.4\textwidth]{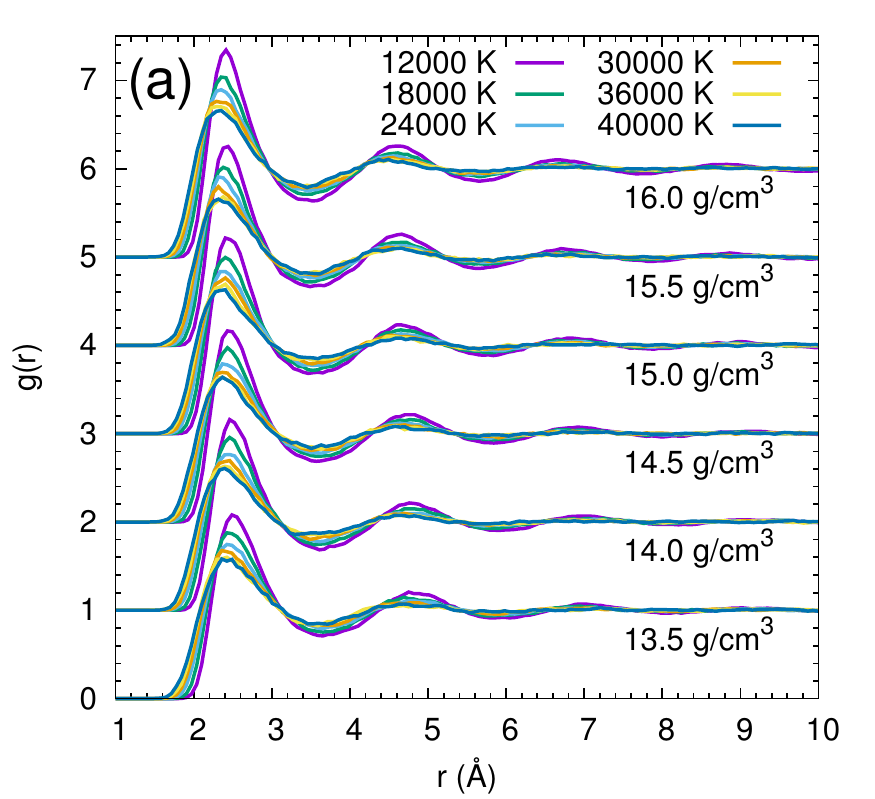}
\includegraphics[width=0.4\textwidth]{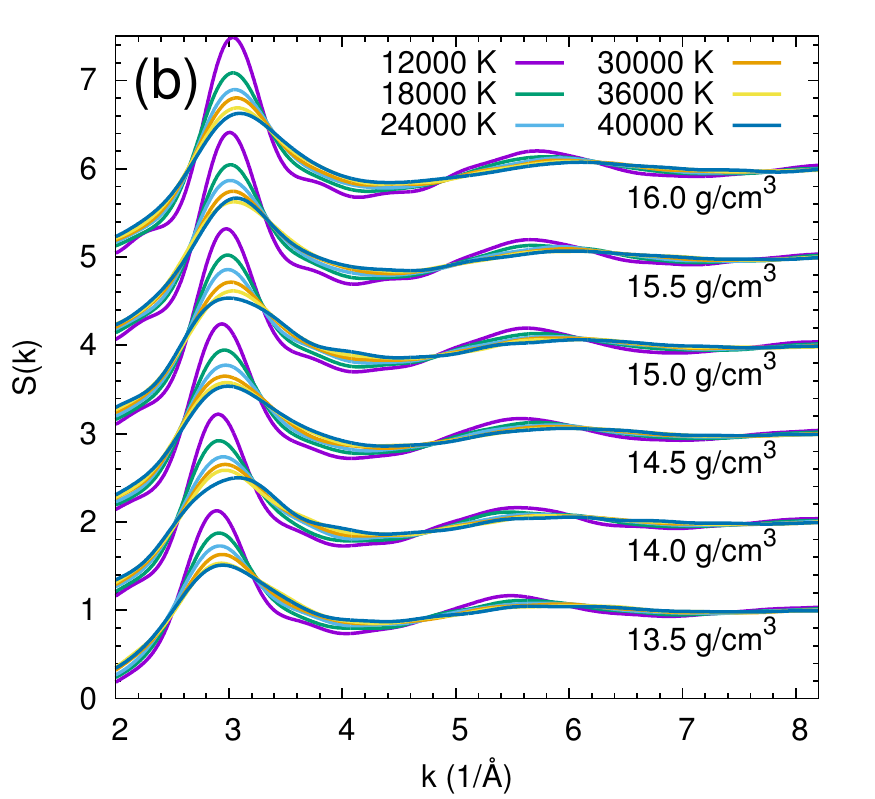} \\ 
\includegraphics[width=0.4\textwidth]{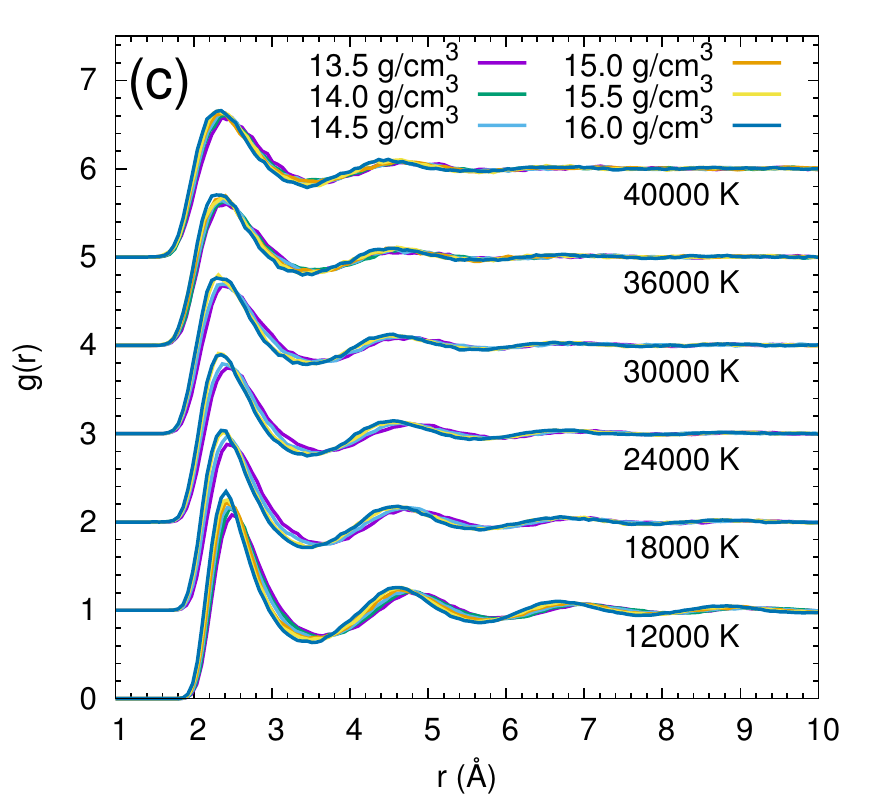}
\includegraphics[width=0.4\textwidth]{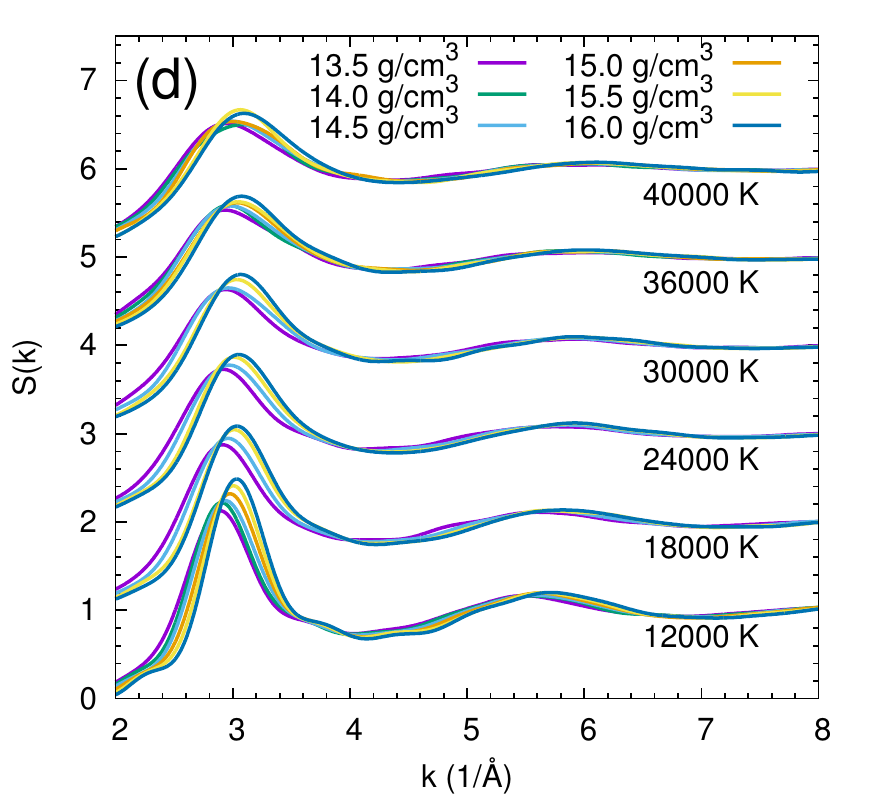}
\caption{\label{fig:strucSn} (left) Pair-correlation function and (right) structure factor of liquid tin at varies temperatures and densities. Different isochores (a-b) and isotherms (c-d) have been shifted apart for clarity.}
\end{figure*}

\begin{figure*}
\includegraphics[width=0.4\textwidth]{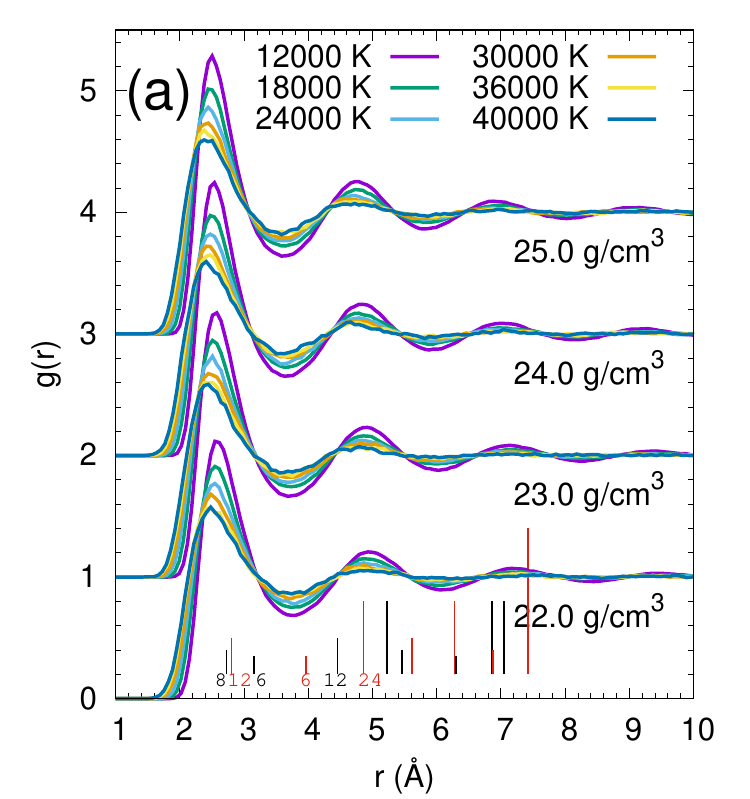}
\includegraphics[width=0.4\textwidth]{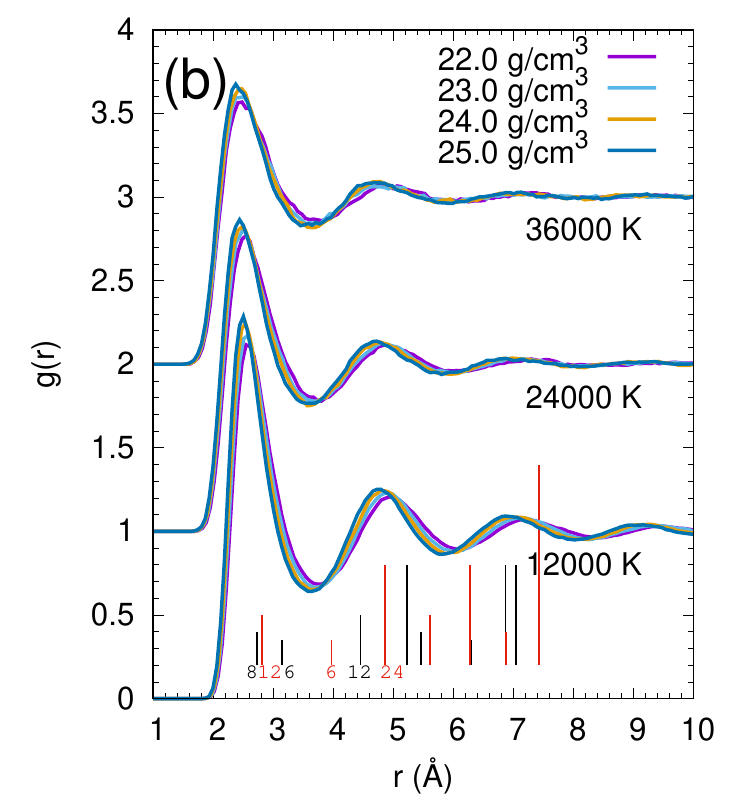}
\caption{\label{fig:strucPb} Pair-correlation function of liquid lead at varies temperatures and densities. Different isochores (a) and isotherms (b) have been shifted apart for clarity. Positions of the up-to-eighth (seventh)-nearest-neighbor shells of a bcc (fcc) crystal at 22.0 g/cm$^3$ are also shown with black (red) vertical bars for comparison. Notations for the bars are the same as Fig.~\ref{fig:strucCu}.}
\end{figure*}

\subsection{\label{subsec:gr}Atomistic structure}

We characterize the atomistic structure using the radial distribution
function $g(r)$, which reflects the local density changes
as a function of the inter-atomic distance.
The $g(r)$ profiles of Cu at selected densities and temperatures are summarized in Fig.~\ref{fig:strucCu}.
At 6000 K, the $g(r)$ results show distinct peak-valley structures and long-range correlations, which are typical for a solid at high temperature
and is consistent with that expected from a bcc crystal. This provides direct evidence for the stability of solid Cu at these conditions and 
is consistent with the EOS discontinuity 
that was discussed in the previous sub-section.
Note that Cu in its ground state remains an fcc structure at pressures as high as 10--23 Mbar~\cite{Fratanduono2020}, whose melting temperature, if estimated to $\sim200$ GPa based on the measurements by Japel {\it et al}.~\cite{Japel2005prl}, 
is in reasonable consistency with our estimated range (6000--7000 K) for the melting temperature of the bcc structure.

At temperatures of 8000 K or higher, the overall profiles of  $g(r)$ are similar among different isochores or isotherms: a primary peak exists at 2~\AA, which is followed by a valley at 3~\AA, two additional peaks at 3.8 and 5.6~\AA\ that become weaker at larger distances, and a smoothly flattened tail at 6.2~\AA\ or greater. The primary peak has a height exceeding 2, and it gradually decreases and broadens as temperature increases, or increases and shifts to smaller $r$ as density increases. Coordination analysis of the liquid states indicates a coordination  number CN=$4\pi n\int_0^{r_\text{valley1}}g(r)r^2dr$=12--14 for the first shell and the value is not clearly dependent on temperature or density, consistent with that of 
a simple close-packed liquid structure. Note that we would get CN=5--7 if using $8\pi n\int_0^{r_\text{peak1}}g(r)r^2dr$, which indicates that, in liquid Cu, the nearest-neighbor atoms are pushed to larger $r$ and the second-nearest-neighbor atoms are pulled to smaller $r$ relative to those in a bcc solid.

Our $g(r)$ profiles for Fe (Fig.~\ref{fig:strucFe}) are similar to those of Cu. The results show a bcc solid feature at 6000~K and 13.5~g/cm$^3$ but are simple close-packed liquid structures at all other conditions shown in the diagram.

The Sn and Pb results are shown in Figs.~\ref{fig:strucSn} and \ref{fig:strucPb}. We also present the structure factor $S(k)$ results for Sn in Fig.~\ref{fig:strucSn}. Similar to the those of liquid Cu and Fe, the $g(r)$ of Sn and Pb also has a primary peak that corresponds to CN=12--14 (when integrating to the first valley) and shows similar trends with density and temperature as those of Cu. However, the height of the primary peak gets smaller than 2, because of the high temperatures, and the peak positions look more similar to the neighbor-shell positions of a fcc than bcc solid. It is noteworthy that these are signatures of simple close-packed metallic liquid and do not depend on the initial structure of the simulation cell---we have compared 128- and 256-atom cells for Sn and found the same $g(r)$ profiles. Recent x-ray diffraction measurements of Sn under shock compression to 90 GPa also show liquid structures that are consistent with the DFT-MD predictions~\cite{Briggs2019}.

A tiny bump is observed in Fig.~\ref{fig:strucSn} along the $S(k)$ profiles of Sn at 12000 K but not at higher $T$. This indicates a gradual structural transition in liquid Sn at $T\le12000$ K. However, we observe no clear difference between the $g(r)$ profiles at 12000 K and those at higher temperatures, nor in the $g(r)$ profiles of liquid Pb at the conditions considered in this study. Whether this is associated with LLP or electronic transition is an interesting question that is beyond the scope of discussion in this work and should be addressed in a future paper.

\begin{figure}
\includegraphics[width=0.5\textwidth]{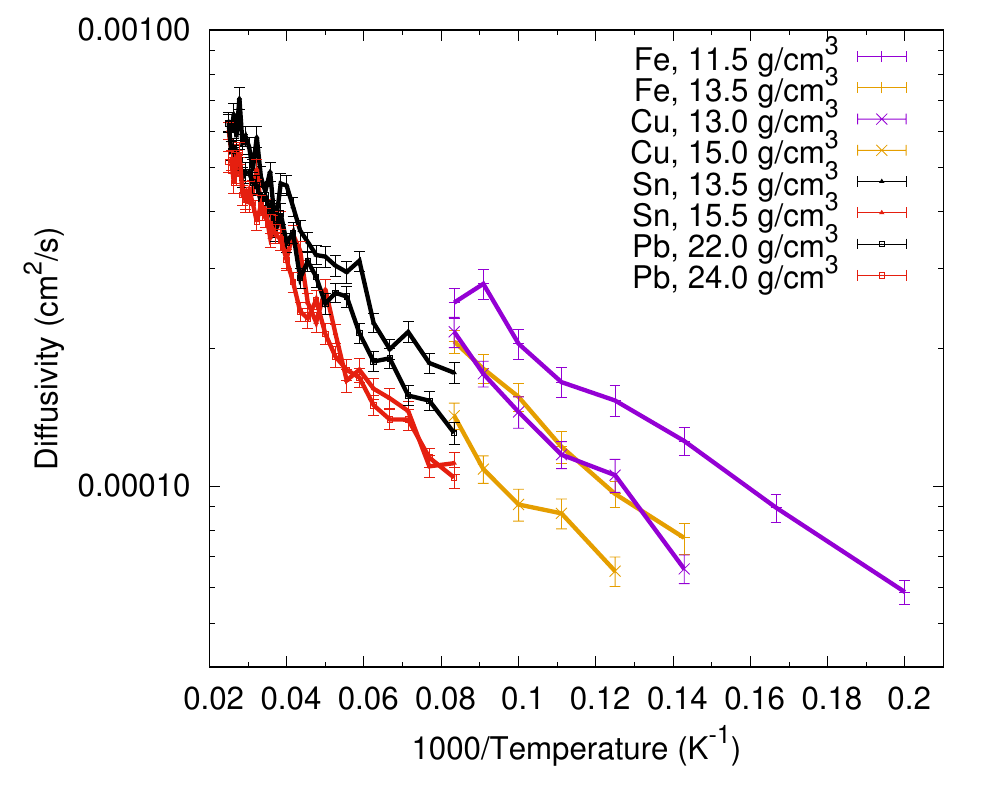}
\caption{\label{fig:diffuse} Arrhenius plot of the self diffusion coefficient for four liquid metals at varies densities.}
\end{figure}

\subsection{\label{subsec:diff}Self diffusivity}

A basic transport property of a liquid is the diffusivity.
This is usually characterized by the Arrhenius relation
$\ln D\propto 1/T$, where $D$ is the self diffusion coefficient and the slop corresponds to an 
``activation energy'' $E_a$, which indicates the energy barrier associated with the packing efficiency of the first
shell of neighboring atoms.

We calculate the self diffusion coefficient $D$ using
the Einstein relation $D=\text{MSD}/6\tau$, where $\tau$ is the length of the simulation in time , and plot the results for the four metals at representative densities 
in Fig.~\ref{fig:diffuse}. Our results show that, for all four metals, diffusion is hindered when density increases or temperature decreases, due to the increased packing or decreased kinetics. The Arrhenius behavior
is satisfied for all the four liquid metals at various
densities. The slope for the different metals 
at different densities are similar, indicating similar
activation energy $E_a$ and coordination environment.
This is consistent with our findings that all four 
metals in their liquid form have simple close-packed
structures, based on the $g(r)$ and CN
analysis in the previous sub-section. 

\section{Conclusion}
In this work, we report first-principles equations of state,
structures, and transport properties of four meltallic systems,
Cu, Fe, Sn, and Pb, near the Hugoniot and at 1.5--5 Mbar
conditions. The calculations are based on state-of-art
quantum molecular dynamics using DFT, which provides an
accurate description of the electronic interactions and
significant advantageous over classical MD approaches.

Our EOS results allow estimating the 
melting temperature of  bcc-Cu to
be 6000--8000 K at 1.5--4 Mbar, similar to that of fcc-Cu
estimated based on measurements at $\le1$ Mbar.
The melting temperature of bcc-Fe is 
$\sim$1000--2000 K lower than that of bcc-Cu
and the Clapeyron slopes are similar,
at variance with previous observations at pressures
lower than 1 Mbar.

Our predictions of the shock Hugoniot benchmark 
the performance of EOS models for these materials.
By comparing our first-principles Hugoniots to predictions of LEOS models, we find the Sn model 
(LEOS 500) is softer by $\sim$10\% but gives reasonable temperature-pressure relation; the Pb 
model (LEOS 820) gives similar pressure-density 
relation but the Hugoniot temperature is lower, more so at 
higher pressures with a maximum of deviation of 10\% at 
4 Mbar; the Fe model (LEOS 260) and the Cu model (LEOS 290) also give similar 
pressure-density relations to our first-principles predictions for the liquid, 
but the Hugoniot temperatures are higher by $\sim$30\%.

Our structural analysis indicates all
the four metals in the liquid state and 1.5--5 Mbar 
are close-packed simple liquids,
as characterized by a maximum peak in $g(r)$ at 2--3~\AA\ 
that decreases both in distance and height with temperature
and decreases in distance but increases in height 
with density, and correspond to a coordination number of 12--14.

The self diffusion coefficient show good Arrhenius behavior with respect to temperature and indicate similar activation energy among 
the four different liquid metals, re-affirming that they form simple close-packed structures at the multi-Mbar conditions.

\section{Appendix}
The equation of state data for Fe, Cu, Pb, and Sn at megabar pressures calculated using DFTMD and reported in this manuscript are listed in the Table~\ref{tab:eostable}.

\begin{acknowledgments}
This work was performed under the auspices of the U.S. Department of Energy by Lawrence Livermore National Laboratory (LLNL) under Contract No. DE-AC52-07NA27344.
We acknowledge support from the LLNL Lab Directed Research and Development (LDRD) program 18-ERD-012.
Computational support was provided by LLNL high-performance computing facility and the Computational Grand Challenge Award.

This document was prepared as an account of work sponsored by an agency of the United States government. Neither the United States government nor Lawrence Livermore National Security, LLC, nor any of their employees makes any warranty, expressed or implied, or assumes any legal liability or responsibility for the accuracy, completeness, or usefulness of any information, apparatus, product, or process disclosed, or represents that its use would not infringe privately owned rights. Reference herein to any specific commercial product, process, or service by trade name, trademark, manufacturer, or otherwise does not necessarily constitute or imply its endorsement, recommendation, or favoring by the United States government or Lawrence Livermore National Security, LLC. The views and opinions of authors expressed herein do not necessarily state or reflect those of the United States government or Lawrence Livermore National Security, LLC, and shall not be used for advertising or product endorsement purposes.
\end{acknowledgments}

\nocite{*}

%
%

\begin{longtable*}{ccccccl}
\caption{\label{tab:eostable} Equations of state for four metals at megabar pressures calculated using DFTMD. The data include density ($\rho$), temperature ($T$), internal energy ($E$) and its standard error ($E_\text{error}$) , pressure ($P$) and its standard error ($P_\text{error}$), and a description with metal type and state [liquid (liq.), solid (sol.), or unstable (sol*)]}\\\hline
$\rho$ & $T$ & $E$ & $E_\text{error}$ & $P$ & $P_\text{error}$ & note \\
(g/cm$^3$) & (K) & (eV/atom) & (eV/atom) & (GPa) & (GPa) &  \\
\hline 
\endfirsthead
\multicolumn{7}{@{}l}{\ldots continued}\\\hline
$\rho$ & $T$ & $E$ & $E_\text{error}$ & $P$ & $P_\text{error}$ & note \\
(g/cm$^3$) & (K) & (eV/atom) & (eV/atom) & (GPa) & (GPa) &  \\\hline
\endhead 
\hline
\multicolumn{7}{r@{}}{continued \ldots}\\
\endfoot
\hline
\endlastfoot
11.50 & 5000.00 & -5.329891 & 0.002587 & 183.4108 & 0.0884 & Fe, liq. \\
11.50 & 6000.00 & -4.920986 & 0.003193 & 194.5701 & 0.1088 & Fe, liq. \\
11.50 & 7000.00 & -4.515845 & 0.005111 & 205.2158 & 0.2033 & Fe, liq. \\
11.50 & 8000.00 & -4.103636 & 0.004204 & 215.7905 & 0.1613 & Fe, liq. \\
11.50 & 9000.00 & -3.701556 & 0.003636 & 225.8689 & 0.1605 & Fe, liq. \\
11.50 & 10000.00 & -3.298742 & 0.005060 & 235.6782 & 0.1913 & Fe, liq. \\
11.50 & 11000.00 & -2.883125 & 0.006070 & 245.7125 & 0.2170 & Fe, liq. \\
11.50 & 12000.00 & -2.481925 & 0.007336 & 254.9459 & 0.3142 & Fe, liq. \\
12.00 & 6000.00 & -4.621283 & 0.005232 & 230.8065 & 0.1959 & Fe, liq. \\
12.00 & 7000.00 & -4.207331 & 0.005723 & 242.0172 & 0.2255 & Fe, liq. \\
12.00 & 8000.00 & -3.795274 & 0.003834 & 252.8390 & 0.1497 & Fe, liq. \\
12.00 & 9000.00 & -3.381674 & 0.004511 & 263.4879 & 0.1996 & Fe, liq. \\
12.00 & 10000.00 & -2.970353 & 0.005461 & 273.8140 & 0.2080 & Fe, liq. \\
12.00 & 11000.00 & -2.564701 & 0.007380 & 283.7254 & 0.2975 & Fe, liq. \\
12.00 & 12000.00 & -2.146736 & 0.004791 & 293.9316 & 0.2175 & Fe, liq. \\
12.50 & 6000.00 & -4.278268 & 0.004147 & 270.7420 & 0.1714 & Fe, liq. \\
12.50 & 7000.00 & -3.851087 & 0.003717 & 282.5037 & 0.1414 & Fe, liq. \\
12.50 & 8000.00 & -3.429263 & 0.003986 & 293.9237 & 0.1489 & Fe, liq. \\
12.50 & 9000.00 & -3.027538 & 0.004318 & 304.3220 & 0.1791 & Fe, liq. \\
12.50 & 10000.00 & -2.609267 & 0.004354 & 315.1017 & 0.1528 & Fe, liq. \\
12.50 & 11000.00 & -2.209792 & 0.006523 & 324.9948 & 0.2362 & Fe, liq. \\
12.50 & 12000.00 & -1.772056 & 0.006011 & 336.1317 & 0.2284 & Fe, liq. \\
13.00 & 7000.00 & -3.446895 & 0.006202 & 326.8339 & 0.2208 & Fe, liq. \\
13.00 & 8000.00 & -3.027860 & 0.004445 & 338.2889 & 0.1530 & Fe, liq. \\
13.00 & 9000.00 & -2.612184 & 0.004794 & 349.2825 & 0.1808 & Fe, liq. \\
13.00 & 10000.00 & -2.197207 & 0.006100 & 360.0585 & 0.2116 & Fe, liq. \\
13.00 & 11000.00 & -1.784707 & 0.008972 & 370.7469 & 0.3188 & Fe, liq. \\
13.00 & 12000.00 & -1.363271 & 0.008094 & 381.3199 & 0.3078 & Fe, liq. \\
13.50 & 7000.00 & -3.014449 & 0.004688 & 374.5179 & 0.1422 & Fe, liq. \\
13.50 & 8000.00 & -2.587004 & 0.004704 & 386.2468 & 0.1585 & Fe, liq. \\
13.50 & 9000.00 & -2.163159 & 0.004989 & 397.6389 & 0.1818 & Fe, liq. \\
13.50 & 10000.00 & -1.741604 & 0.005778 & 408.7730 & 0.1728 & Fe, liq. \\
13.50 & 11000.00 & -1.322616 & 0.005352 & 419.7924 & 0.1904 & Fe, liq. \\
13.50 & 12000.00 & -0.910728 & 0.009000 & 430.1242 & 0.3318 & Fe, liq. \\
13.00 & 7000.00 & -0.094671 & 0.005753 & 219.5882 & 0.2648 & Cu, liq. \\
13.00 & 8000.00 & 0.250756 & 0.004860 & 228.5770 & 0.2256 & Cu, liq. \\
13.00 & 9000.00 & 0.604556 & 0.011443 & 237.6012 & 0.5142 & Cu, liq. \\
13.00 & 10000.00 & 0.953658 & 0.007691 & 246.0420 & 0.3147 & Cu, liq. \\
13.00 & 11000.00 & 1.308474 & 0.011894 & 254.1927 & 0.5089 & Cu, liq. \\
13.00 & 12000.00 & 1.678023 & 0.010215 & 262.9623 & 0.4646 & Cu, liq. \\
13.50 & 8000.00 & 0.593897 & 0.007870 & 262.2252 & 0.3600 & Cu, liq. \\
13.50 & 9000.00 & 0.907804 & 0.011270 & 269.7825 & 0.4953 & Cu, liq. \\
13.50 & 10000.00 & 1.275812 & 0.009172 & 279.2658 & 0.4176 & Cu, liq. \\
13.50 & 11000.00 & 1.643082 & 0.007316 & 288.5774 & 0.3325 & Cu, liq. \\
13.50 & 12000.00 & 2.006104 & 0.007170 & 297.0766 & 0.3059 & Cu, liq. \\
14.00 & 7000.00 & 0.555745 & 0.012347 & 286.2523 & 0.5951 & Cu, liq. \\
14.00 & 8000.00 & 0.931516 & 0.008220 & 297.2385 & 0.3800 & Cu, liq. \\
14.00 & 9000.00 & 1.287103 & 0.006000 & 306.9231 & 0.2912 & Cu, liq. \\
14.00 & 10000.00 & 1.619758 & 0.006150 & 315.1948 & 0.2592 & Cu, liq. \\
14.00 & 11000.00 & 1.994450 & 0.004867 & 325.1881 & 0.2511 & Cu, liq. \\
14.00 & 12000.00 & 2.362595 & 0.008414 & 334.3163 & 0.3084 & Cu, liq. \\
14.50 & 8000.00 & 1.323379 & 0.006499 & 336.3662 & 0.2654 & Cu, liq. \\
14.50 & 9000.00 & 1.652881 & 0.009508 & 345.2224 & 0.4266 & Cu, liq. \\
14.50 & 10000.00 & 2.033146 & 0.008177 & 356.0007 & 0.3656 & Cu, liq. \\
14.50 & 11000.00 & 2.369111 & 0.014834 & 364.4996 & 0.7063 & Cu, liq. \\
14.50 & 12000.00 & 2.767980 & 0.014097 & 375.3226 & 0.5558 & Cu, liq. \\
15.00 & 8000.00 & 1.737100 & 0.009891 & 378.2839 & 0.4569 & Cu, liq. \\
15.00 & 9000.00 & 2.079838 & 0.008010 & 388.0854 & 0.3715 & Cu, liq. \\
15.00 & 10000.00 & 2.461830 & 0.008849 & 399.1470 & 0.4100 & Cu, liq. \\
15.00 & 11000.00 & 2.808712 & 0.012853 & 408.1596 & 0.5929 & Cu, liq. \\
15.00 & 12000.00 & 3.162047 & 0.007711 & 417.4906 & 0.3758 & Cu, liq. \\
21.50 & 12000.00 & 4.257651 & 0.006685 & 202.9974 & 0.1381 & Pb, liq. \\
21.50 & 13000.00 & 4.593085 & 0.005323 & 206.4020 & 0.1151 & Pb, liq. \\
21.50 & 14000.00 & 4.945054 & 0.005929 & 209.9948 & 0.1205 & Pb, liq. \\
21.50 & 15000.00 & 5.285054 & 0.009325 & 213.1366 & 0.1966 & Pb, liq. \\
21.50 & 16000.00 & 5.660475 & 0.006324 & 216.8797 & 0.1307 & Pb, liq. \\
21.50 & 17000.00 & 6.029187 & 0.010200 & 220.2663 & 0.2099 & Pb, liq. \\
21.50 & 18000.00 & 6.401480 & 0.009620 & 223.5672 & 0.1963 & Pb, liq. \\
21.50 & 19000.00 & 6.784173 & 0.011451 & 226.8480 & 0.2311 & Pb, liq. \\
21.50 & 20000.00 & 7.206125 & 0.012264 & 230.7629 & 0.2415 & Pb, liq. \\
21.50 & 21000.00 & 7.591985 & 0.014902 & 233.8311 & 0.3090 & Pb, liq. \\
21.50 & 22000.00 & 8.016126 & 0.014176 & 237.4104 & 0.2885 & Pb, liq. \\
21.50 & 23000.00 & 8.427134 & 0.013593 & 240.5180 & 0.2667 & Pb, liq. \\
21.50 & 24000.00 & 8.858081 & 0.016250 & 243.8413 & 0.3218 & Pb, liq. \\
21.50 & 25000.00 & 9.296254 & 0.015371 & 247.1688 & 0.2807 & Pb, liq. \\
21.50 & 26000.00 & 9.735717 & 0.014271 & 250.3447 & 0.2676 & Pb, liq. \\
21.50 & 27000.00 & 10.188634 & 0.017904 & 253.5865 & 0.3391 & Pb, liq. \\
21.50 & 28000.00 & 10.617116 & 0.017677 & 256.1325 & 0.3464 & Pb, liq. \\
21.50 & 29000.00 & 11.135861 & 0.011727 & 260.2429 & 0.2414 & Pb, liq. \\
21.50 & 30000.00 & 11.575125 & 0.022351 & 262.7635 & 0.3809 & Pb, liq. \\
21.50 & 31000.00 & 12.111450 & 0.021588 & 266.8734 & 0.3819 & Pb, liq. \\
21.50 & 32000.00 & 12.559619 & 0.014196 & 269.1354 & 0.2701 & Pb, liq. \\
21.50 & 33000.00 & 13.095212 & 0.031285 & 272.9958 & 0.5719 & Pb, liq. \\
21.50 & 34000.00 & 13.534431 & 0.036653 & 274.7562 & 0.6554 & Pb, liq. \\
21.50 & 35000.00 & 14.121419 & 0.026019 & 279.1447 & 0.4353 & Pb, liq. \\
21.50 & 36000.00 & 14.691167 & 0.014942 & 282.9355 & 0.2412 & Pb, liq. \\
21.50 & 37000.00 & 15.173474 & 0.020574 & 285.0601 & 0.3205 & Pb, liq. \\
21.50 & 38000.00 & 15.755867 & 0.022269 & 289.1426 & 0.4191 & Pb, liq. \\
21.50 & 39000.00 & 16.244750 & 0.031502 & 290.9869 & 0.5099 & Pb, liq. \\
21.50 & 40000.00 & 16.865688 & 0.037695 & 295.5124 & 0.6132 & Pb, liq. \\
22.00 & 12000.00 & 4.611033 & 0.009053 & 217.7433 & 0.1926 & Pb, liq. \\
22.00 & 13000.00 & 4.978210 & 0.011388 & 221.8917 & 0.2391 & Pb, liq. \\
22.00 & 14000.00 & 5.318470 & 0.006679 & 225.2655 & 0.1482 & Pb, liq. \\
22.00 & 15000.00 & 5.680725 & 0.005907 & 228.9417 & 0.1195 & Pb, liq. \\
22.00 & 16000.00 & 6.053077 & 0.007369 & 232.6261 & 0.1521 & Pb, liq. \\
22.00 & 17000.00 & 6.415140 & 0.010300 & 235.9268 & 0.2134 & Pb, liq. \\
22.00 & 18000.00 & 6.783079 & 0.006243 & 239.1802 & 0.1449 & Pb, liq. \\
22.00 & 19000.00 & 7.167708 & 0.016128 & 242.5790 & 0.3350 & Pb, liq. \\
22.00 & 20000.00 & 7.586762 & 0.010802 & 246.4759 & 0.2123 & Pb, liq. \\
22.00 & 21000.00 & 7.968525 & 0.015040 & 249.4381 & 0.3062 & Pb, liq. \\
22.00 & 22000.00 & 8.383054 & 0.017405 & 252.8915 & 0.3474 & Pb, liq. \\
22.00 & 23000.00 & 8.861280 & 0.014526 & 257.4086 & 0.2816 & Pb, liq. \\
22.00 & 24000.00 & 9.204072 & 0.015281 & 259.0350 & 0.2979 & Pb, liq. \\
22.00 & 25000.00 & 9.673726 & 0.013113 & 262.9418 & 0.2562 & Pb, liq. \\
22.00 & 26000.00 & 10.069328 & 0.014764 & 265.2739 & 0.2629 & Pb, liq. \\
22.00 & 27000.00 & 10.573473 & 0.026614 & 269.6616 & 0.5228 & Pb, liq. \\
22.00 & 28000.00 & 11.029983 & 0.022702 & 272.7825 & 0.4140 & Pb, liq. \\
22.00 & 29000.00 & 11.517111 & 0.018485 & 276.4197 & 0.3382 & Pb, liq. \\
22.00 & 30000.00 & 11.987782 & 0.020226 & 279.3949 & 0.3193 & Pb, liq. \\
22.00 & 31000.00 & 12.487005 & 0.030453 & 282.8386 & 0.5397 & Pb, liq. \\
22.00 & 32000.00 & 13.005971 & 0.017042 & 286.6330 & 0.3123 & Pb, liq. \\
22.00 & 33000.00 & 13.537500 & 0.023877 & 290.2181 & 0.4339 & Pb, liq. \\
22.00 & 34000.00 & 13.966346 & 0.034062 & 292.0267 & 0.6287 & Pb, liq. \\
22.00 & 35000.00 & 14.534687 & 0.029837 & 296.1972 & 0.5384 & Pb, liq. \\
22.00 & 36000.00 & 15.103571 & 0.018071 & 299.9891 & 0.3192 & Pb, liq. \\
22.00 & 37000.00 & 15.625405 & 0.032393 & 303.0878 & 0.5997 & Pb, liq. \\
22.00 & 38000.00 & 16.105066 & 0.041879 & 304.9725 & 0.6976 & Pb, liq. \\
22.00 & 39000.00 & 16.596029 & 0.033239 & 307.1166 & 0.6096 & Pb, liq. \\
22.00 & 40000.00 & 17.224267 & 0.031250 & 311.4705 & 0.4517 & Pb, liq. \\
22.50 & 12000.00 & 5.006481 & 0.010033 & 233.8160 & 0.2146 & Pb, liq. \\
22.50 & 13000.00 & 5.358934 & 0.004834 & 237.6836 & 0.1053 & Pb, liq. \\
22.50 & 14000.00 & 5.727071 & 0.009091 & 241.7219 & 0.1994 & Pb, liq. \\
22.50 & 15000.00 & 6.073118 & 0.007303 & 245.0803 & 0.1567 & Pb, liq. \\
22.50 & 16000.00 & 6.435943 & 0.007845 & 248.6721 & 0.1631 & Pb, liq. \\
22.50 & 17000.00 & 6.816889 & 0.011017 & 252.3922 & 0.2195 & Pb, liq. \\
22.50 & 18000.00 & 7.212443 & 0.009612 & 256.2041 & 0.2135 & Pb, liq. \\
22.50 & 19000.00 & 7.590197 & 0.011030 & 259.4787 & 0.2255 & Pb, liq. \\
22.50 & 20000.00 & 7.999462 & 0.011043 & 263.2201 & 0.2216 & Pb, liq. \\
22.50 & 21000.00 & 8.367667 & 0.013200 & 266.0041 & 0.2820 & Pb, liq. \\
22.50 & 22000.00 & 8.760112 & 0.010412 & 269.0053 & 0.2191 & Pb, liq. \\
22.50 & 23000.00 & 9.212213 & 0.017192 & 273.0462 & 0.3505 & Pb, liq. \\
22.50 & 24000.00 & 9.641598 & 0.015722 & 276.4077 & 0.3094 & Pb, liq. \\
22.50 & 25000.00 & 10.094143 & 0.021065 & 280.0975 & 0.4153 & Pb, liq. \\
22.50 & 26000.00 & 10.532394 & 0.013607 & 283.2828 & 0.2493 & Pb, liq. \\
22.50 & 27000.00 & 11.018215 & 0.022790 & 287.1011 & 0.4301 & Pb, liq. \\
22.50 & 28000.00 & 11.408367 & 0.015986 & 289.1972 & 0.3036 & Pb, liq. \\
22.50 & 29000.00 & 11.942074 & 0.024099 & 293.6437 & 0.4643 & Pb, liq. \\
22.50 & 30000.00 & 12.361293 & 0.021590 & 295.8724 & 0.4140 & Pb, liq. \\
22.50 & 31000.00 & 12.838036 & 0.018176 & 298.9626 & 0.3371 & Pb, liq. \\
22.50 & 32000.00 & 13.472163 & 0.024795 & 304.5149 & 0.4305 & Pb, liq. \\
22.50 & 33000.00 & 13.890761 & 0.023800 & 306.3293 & 0.3991 & Pb, liq. \\
22.50 & 34000.00 & 14.433232 & 0.030327 & 310.1224 & 0.5748 & Pb, liq. \\
22.50 & 35000.00 & 14.916572 & 0.036915 & 312.7340 & 0.6825 & Pb, liq. \\
22.50 & 36000.00 & 15.404848 & 0.024736 & 315.1531 & 0.4267 & Pb, liq. \\
22.50 & 37000.00 & 16.000122 & 0.042736 & 319.5569 & 0.7313 & Pb, liq. \\
22.50 & 38000.00 & 16.547979 & 0.027121 & 322.7578 & 0.4654 & Pb, liq. \\
22.50 & 39000.00 & 17.054990 & 0.053969 & 325.2672 & 0.8815 & Pb, liq. \\
22.50 & 40000.00 & 17.639297 & 0.049391 & 329.0298 & 0.8247 & Pb, liq. \\
23.00 & 12000.00 & 5.398889 & 0.007668 & 250.2825 & 0.1693 & Pb, liq. \\
23.00 & 13000.00 & 5.771459 & 0.008180 & 254.6444 & 0.1748 & Pb, liq. \\
23.00 & 14000.00 & 6.113699 & 0.010408 & 258.1522 & 0.2249 & Pb, liq. \\
23.00 & 15000.00 & 6.478070 & 0.009076 & 261.9664 & 0.2011 & Pb, liq. \\
23.00 & 16000.00 & 6.883011 & 0.010785 & 266.4058 & 0.2361 & Pb, liq. \\
23.00 & 17000.00 & 7.208523 & 0.009921 & 269.0117 & 0.2127 & Pb, liq. \\
23.00 & 18000.00 & 7.605003 & 0.014604 & 272.9327 & 0.3070 & Pb, liq. \\
23.00 & 19000.00 & 8.009378 & 0.014481 & 276.7311 & 0.2931 & Pb, liq. \\
23.00 & 20000.00 & 8.372145 & 0.018067 & 279.5946 & 0.3853 & Pb, liq. \\
23.00 & 21000.00 & 8.811422 & 0.014834 & 283.7417 & 0.2960 & Pb, liq. \\
23.00 & 22000.00 & 9.222150 & 0.010855 & 287.2735 & 0.2074 & Pb, liq. \\
23.00 & 23000.00 & 9.660186 & 0.015945 & 291.0011 & 0.3206 & Pb, liq. \\
23.00 & 24000.00 & 10.095103 & 0.025938 & 294.5880 & 0.5160 & Pb, liq. \\
23.00 & 25000.00 & 10.492560 & 0.011397 & 297.1618 & 0.2385 & Pb, liq. \\
23.00 & 26000.00 & 10.980228 & 0.013069 & 301.3481 & 0.2724 & Pb, liq. \\
23.00 & 27000.00 & 11.400238 & 0.012154 & 304.0059 & 0.2385 & Pb, liq. \\
23.00 & 28000.00 & 11.849102 & 0.031724 & 307.1553 & 0.6097 & Pb, liq. \\
23.00 & 29000.00 & 12.386978 & 0.026667 & 311.8047 & 0.4909 & Pb, liq. \\
23.00 & 30000.00 & 12.834747 & 0.026109 & 314.4806 & 0.5123 & Pb, liq. \\
23.00 & 31000.00 & 13.331368 & 0.030306 & 317.9744 & 0.5252 & Pb, liq. \\
23.00 & 32000.00 & 13.811607 & 0.029431 & 321.0224 & 0.5352 & Pb, liq. \\
23.00 & 33000.00 & 14.346605 & 0.021720 & 324.6748 & 0.4092 & Pb, liq. \\
23.00 & 34000.00 & 14.883307 & 0.048249 & 328.3880 & 0.8764 & Pb, liq. \\
23.00 & 35000.00 & 15.402465 & 0.027038 & 331.7019 & 0.4878 & Pb, liq. \\
23.00 & 36000.00 & 15.947679 & 0.018469 & 335.2384 & 0.3709 & Pb, liq. \\
23.00 & 37000.00 & 16.418044 & 0.034460 & 337.4895 & 0.6233 & Pb, liq. \\
23.00 & 38000.00 & 17.050227 & 0.031156 & 342.1625 & 0.5614 & Pb, liq. \\
23.00 & 39000.00 & 17.591484 & 0.041252 & 345.1954 & 0.6635 & Pb, liq. \\
23.00 & 40000.00 & 18.093782 & 0.038055 & 347.7126 & 0.7025 & Pb, liq. \\
23.50 & 12000.00 & 5.832144 & 0.005531 & 268.0944 & 0.1254 & Pb, liq. \\
23.50 & 13000.00 & 6.180798 & 0.006871 & 271.9304 & 0.1588 & Pb, liq. \\
23.50 & 14000.00 & 6.538166 & 0.008161 & 275.8477 & 0.1774 & Pb, liq. \\
23.50 & 15000.00 & 6.909544 & 0.007755 & 279.7826 & 0.1725 & Pb, liq. \\
23.50 & 16000.00 & 7.290022 & 0.009572 & 283.8429 & 0.1983 & Pb, liq. \\
23.50 & 17000.00 & 7.665382 & 0.008759 & 287.4973 & 0.1953 & Pb, liq. \\
23.50 & 18000.00 & 8.031079 & 0.011217 & 290.7358 & 0.2367 & Pb, liq. \\
23.50 & 19000.00 & 8.430870 & 0.011508 & 294.6016 & 0.2473 & Pb, liq. \\
23.50 & 20000.00 & 8.819532 & 0.009067 & 297.9767 & 0.2013 & Pb, liq. \\
23.50 & 21000.00 & 9.249486 & 0.014700 & 302.0242 & 0.3059 & Pb, liq. \\
23.50 & 22000.00 & 9.637950 & 0.021856 & 304.9350 & 0.4447 & Pb, liq. \\
23.50 & 23000.00 & 10.081183 & 0.012460 & 308.8989 & 0.2601 & Pb, liq. \\
23.50 & 24000.00 & 10.491432 & 0.018280 & 311.9396 & 0.3935 & Pb, liq. \\
23.50 & 25000.00 & 10.954583 & 0.020738 & 315.8305 & 0.4058 & Pb, liq. \\
23.50 & 26000.00 & 11.385390 & 0.022724 & 318.9958 & 0.4348 & Pb, liq. \\
23.50 & 27000.00 & 11.863082 & 0.021520 & 322.8793 & 0.4096 & Pb, liq. \\
23.50 & 28000.00 & 12.361370 & 0.020494 & 326.7929 & 0.4111 & Pb, liq. \\
23.50 & 29000.00 & 12.824273 & 0.023357 & 330.0564 & 0.4726 & Pb, liq. \\
23.50 & 30000.00 & 13.331946 & 0.020190 & 333.9911 & 0.3730 & Pb, liq. \\
23.50 & 31000.00 & 13.739021 & 0.018793 & 335.6686 & 0.3464 & Pb, liq. \\
23.50 & 32000.00 & 14.214167 & 0.025032 & 338.6296 & 0.4839 & Pb, liq. \\
23.50 & 33000.00 & 14.762019 & 0.024041 & 342.6939 & 0.4621 & Pb, liq. \\
23.50 & 34000.00 & 15.289450 & 0.030857 & 346.4947 & 0.5651 & Pb, liq. \\
23.50 & 35000.00 & 15.781956 & 0.009450 & 348.9637 & 0.1623 & Pb, liq. \\
23.50 & 36000.00 & 16.273090 & 0.038631 & 351.7942 & 0.6428 & Pb, liq. \\
23.50 & 37000.00 & 16.923880 & 0.037706 & 357.0699 & 0.6964 & Pb, liq. \\
23.50 & 38000.00 & 17.362780 & 0.039223 & 358.8745 & 0.7262 & Pb, liq. \\
23.50 & 39000.00 & 17.994660 & 0.034341 & 363.4945 & 0.6068 & Pb, liq. \\
23.50 & 40000.00 & 18.487140 & 0.035423 & 365.6236 & 0.6121 & Pb, liq. \\
24.00 & 12000.00 & 6.252190 & 0.004265 & 286.0798 & 0.0969 & Pb, liq. \\
24.00 & 13000.00 & 6.638542 & 0.007437 & 290.8181 & 0.1682 & Pb, liq. \\
24.00 & 14000.00 & 6.978849 & 0.009991 & 294.3676 & 0.2224 & Pb, liq. \\
24.00 & 15000.00 & 7.362684 & 0.013450 & 298.5988 & 0.2846 & Pb, liq. \\
24.00 & 16000.00 & 7.703163 & 0.009122 & 301.7878 & 0.1938 & Pb, liq. \\
24.00 & 17000.00 & 8.105230 & 0.008967 & 306.0561 & 0.2058 & Pb, liq. \\
24.00 & 18000.00 & 8.477045 & 0.011451 & 309.4686 & 0.2499 & Pb, liq. \\
24.00 & 19000.00 & 8.872989 & 0.012443 & 313.2179 & 0.2583 & Pb, liq. \\
24.00 & 20000.00 & 9.262678 & 0.017026 & 316.7121 & 0.3609 & Pb, liq. \\
24.00 & 21000.00 & 9.684660 & 0.010998 & 320.6182 & 0.2385 & Pb, liq. \\
24.00 & 22000.00 & 10.096159 & 0.015757 & 324.0527 & 0.3334 & Pb, liq. \\
24.00 & 23000.00 & 10.568438 & 0.015721 & 328.5545 & 0.3075 & Pb, liq. \\
24.00 & 24000.00 & 10.953062 & 0.017337 & 331.1240 & 0.3506 & Pb, liq. \\
24.00 & 25000.00 & 11.394550 & 0.022082 & 334.6163 & 0.4446 & Pb, liq. \\
24.00 & 26000.00 & 11.867845 & 0.026151 & 338.5821 & 0.5265 & Pb, liq. \\
24.00 & 27000.00 & 12.303087 & 0.028469 & 341.6719 & 0.5782 & Pb, liq. \\
24.00 & 28000.00 & 12.735801 & 0.027137 & 344.4556 & 0.5263 & Pb, liq. \\
24.00 & 29000.00 & 13.214548 & 0.026134 & 347.9928 & 0.5026 & Pb, liq. \\
24.00 & 30000.00 & 13.724821 & 0.037978 & 351.9581 & 0.7332 & Pb, liq. \\
24.00 & 31000.00 & 14.192517 & 0.029761 & 354.7900 & 0.5702 & Pb, liq. \\
24.00 & 32000.00 & 14.723966 & 0.029439 & 358.8342 & 0.5753 & Pb, liq. \\
24.00 & 33000.00 & 15.176981 & 0.021530 & 361.2086 & 0.3713 & Pb, liq. \\
24.00 & 34000.00 & 15.754196 & 0.029879 & 365.7724 & 0.5656 & Pb, liq. \\
24.00 & 35000.00 & 16.288647 & 0.017777 & 369.1717 & 0.3546 & Pb, liq. \\
24.00 & 36000.00 & 16.733197 & 0.029029 & 371.1834 & 0.4948 & Pb, liq. \\
24.00 & 37000.00 & 17.267725 & 0.036622 & 374.4690 & 0.6659 & Pb, liq. \\
24.00 & 38000.00 & 17.875098 & 0.037422 & 378.8908 & 0.6571 & Pb, liq. \\
24.00 & 39000.00 & 18.450022 & 0.040227 & 382.4296 & 0.6126 & Pb, liq. \\
24.00 & 40000.00 & 19.070812 & 0.051382 & 386.8146 & 0.8438 & Pb, liq. \\
24.50 & 12000.00 & 6.719278 & 0.007295 & 305.5160 & 0.1544 & Pb, liq. \\
24.50 & 13000.00 & 7.092026 & 0.008428 & 310.0063 & 0.1912 & Pb, liq. \\
24.50 & 14000.00 & 7.424676 & 0.010340 & 313.4049 & 0.2294 & Pb, liq. \\
24.50 & 15000.00 & 7.812095 & 0.009309 & 317.8076 & 0.2108 & Pb, liq. \\
24.50 & 16000.00 & 8.161333 & 0.009410 & 321.1600 & 0.2081 & Pb, liq. \\
24.50 & 17000.00 & 8.565689 & 0.008074 & 325.4994 & 0.1758 & Pb, liq. \\
24.50 & 18000.00 & 8.948216 & 0.010297 & 329.1919 & 0.2178 & Pb, liq. \\
24.50 & 19000.00 & 9.356696 & 0.012924 & 333.2328 & 0.2675 & Pb, liq. \\
24.50 & 20000.00 & 9.717373 & 0.010895 & 336.0399 & 0.2434 & Pb, liq. \\
24.50 & 21000.00 & 10.123907 & 0.012856 & 339.6522 & 0.2742 & Pb, liq. \\
24.50 & 22000.00 & 10.563665 & 0.014524 & 343.7016 & 0.3111 & Pb, liq. \\
24.50 & 23000.00 & 10.977978 & 0.026835 & 347.0965 & 0.5430 & Pb, liq. \\
24.50 & 24000.00 & 11.438500 & 0.018847 & 351.1038 & 0.4027 & Pb, liq. \\
24.50 & 25000.00 & 11.864180 & 0.016290 & 354.3838 & 0.3367 & Pb, liq. \\
24.50 & 26000.00 & 12.316150 & 0.017091 & 357.9906 & 0.3467 & Pb, liq. \\
24.50 & 27000.00 & 12.803266 & 0.023266 & 362.0321 & 0.4734 & Pb, liq. \\
24.50 & 28000.00 & 13.229098 & 0.033116 & 364.7328 & 0.6705 & Pb, liq. \\
24.50 & 29000.00 & 13.744248 & 0.018943 & 368.9204 & 0.3439 & Pb, liq. \\
24.50 & 30000.00 & 14.232734 & 0.033131 & 372.4987 & 0.6298 & Pb, liq. \\
24.50 & 31000.00 & 14.658654 & 0.029925 & 374.6456 & 0.5526 & Pb, liq. \\
24.50 & 32000.00 & 15.205203 & 0.032315 & 378.7812 & 0.6225 & Pb, liq. \\
24.50 & 33000.00 & 15.653630 & 0.036436 & 381.0634 & 0.6804 & Pb, liq. \\
24.50 & 34000.00 & 16.141745 & 0.033100 & 384.0961 & 0.6372 & Pb, liq. \\
24.50 & 35000.00 & 16.735495 & 0.033580 & 388.8458 & 0.6250 & Pb, liq. \\
24.50 & 36000.00 & 17.241169 & 0.028554 & 391.7941 & 0.5340 & Pb, liq. \\
24.50 & 37000.00 & 17.793055 & 0.035598 & 395.2445 & 0.6415 & Pb, liq. \\
24.50 & 38000.00 & 18.277036 & 0.036869 & 397.6534 & 0.6424 & Pb, liq. \\
24.50 & 39000.00 & 18.825018 & 0.029831 & 401.0589 & 0.5223 & Pb, liq. \\
24.50 & 40000.00 & 19.555848 & 0.048330 & 407.1940 & 0.8745 & Pb, liq. \\
25.00 & 12000.00 & 7.190824 & 0.010384 & 325.6019 & 0.2310 & Pb, liq. \\
25.00 & 13000.00 & 7.558091 & 0.007640 & 329.9873 & 0.1721 & Pb, liq. \\
25.00 & 14000.00 & 7.908311 & 0.006665 & 333.7750 & 0.1580 & Pb, liq. \\
25.00 & 15000.00 & 8.263016 & 0.008336 & 337.4595 & 0.1842 & Pb, liq. \\
25.00 & 16000.00 & 8.646922 & 0.007200 & 341.6022 & 0.1623 & Pb, liq. \\
25.00 & 17000.00 & 9.034660 & 0.010107 & 345.6267 & 0.2155 & Pb, liq. \\
25.00 & 18000.00 & 9.426087 & 0.015190 & 349.5497 & 0.3210 & Pb, liq. \\
25.00 & 19000.00 & 9.815375 & 0.013985 & 353.0994 & 0.2923 & Pb, liq. \\
25.00 & 20000.00 & 10.248660 & 0.009609 & 357.5820 & 0.2045 & Pb, liq. \\
25.00 & 21000.00 & 10.614297 & 0.017890 & 360.2937 & 0.3750 & Pb, liq. \\
25.00 & 22000.00 & 11.050021 & 0.011951 & 364.2386 & 0.2612 & Pb, liq. \\
25.00 & 23000.00 & 11.472756 & 0.016128 & 367.7255 & 0.3217 & Pb, liq. \\
25.00 & 24000.00 & 11.922769 & 0.020282 & 371.6892 & 0.4373 & Pb, liq. \\
25.00 & 25000.00 & 12.341566 & 0.022126 & 374.7731 & 0.4495 & Pb, liq. \\
25.00 & 26000.00 & 12.807641 & 0.018845 & 378.6436 & 0.3810 & Pb, liq. \\
25.00 & 27000.00 & 13.230108 & 0.017447 & 381.4993 & 0.3610 & Pb, liq. \\
25.00 & 28000.00 & 13.716568 & 0.026682 & 385.2313 & 0.4699 & Pb, liq. \\
25.00 & 29000.00 & 14.193074 & 0.021583 & 388.6847 & 0.4239 & Pb, liq. \\
25.00 & 30000.00 & 14.695771 & 0.029721 & 392.7578 & 0.5822 & Pb, liq. \\
25.00 & 31000.00 & 15.217316 & 0.025881 & 396.5626 & 0.4945 & Pb, liq. \\
25.00 & 32000.00 & 15.673535 & 0.015284 & 399.1824 & 0.3091 & Pb, liq. \\
25.00 & 33000.00 & 16.170282 & 0.034556 & 402.4882 & 0.5952 & Pb, liq. \\
25.00 & 34000.00 & 16.719597 & 0.026103 & 406.5938 & 0.5091 & Pb, liq. \\
25.00 & 35000.00 & 17.270099 & 0.045282 & 410.3401 & 0.7427 & Pb, liq. \\
25.00 & 36000.00 & 17.749282 & 0.023478 & 412.7131 & 0.4413 & Pb, liq. \\
25.00 & 37000.00 & 18.339882 & 0.039472 & 416.9052 & 0.6760 & Pb, liq. \\
25.00 & 38000.00 & 18.865165 & 0.031569 & 419.8781 & 0.5475 & Pb, liq. \\
25.00 & 39000.00 & 19.415630 & 0.029216 & 422.9778 & 0.4946 & Pb, liq. \\
25.00 & 40000.00 & 19.937413 & 0.050677 & 426.1921 & 0.7729 & Pb, liq. \\
13.50 & 12000.00 & 3.460350 & 0.009666 & 209.4937 & 0.2492 & Sn, liq. \\
13.50 & 13000.00 & 3.783102 & 0.013804 & 213.1680 & 0.3498 & Sn, liq. \\
13.50 & 14000.00 & 4.102613 & 0.019125 & 216.6174 & 0.4696 & Sn, liq. \\
13.50 & 15000.00 & 4.493744 & 0.021180 & 221.6782 & 0.5389 & Sn, liq. \\
13.50 & 16000.00 & 4.817294 & 0.019328 & 224.8442 & 0.4661 & Sn, liq. \\
13.50 & 17000.00 & 5.209212 & 0.019538 & 229.5568 & 0.4929 & Sn, liq. \\
13.50 & 18000.00 & 5.569239 & 0.015536 & 233.3277 & 0.3840 & Sn, liq. \\
13.50 & 19000.00 & 5.952242 & 0.020897 & 237.3865 & 0.5110 & Sn, liq. \\
13.50 & 20000.00 & 6.355108 & 0.024313 & 241.6333 & 0.5232 & Sn, liq. \\
13.50 & 21000.00 & 6.736175 & 0.013541 & 245.5174 & 0.3389 & Sn, liq. \\
13.50 & 22000.00 & 7.077196 & 0.022843 & 248.0733 & 0.5275 & Sn, liq. \\
13.50 & 23000.00 & 7.494393 & 0.013946 & 252.3143 & 0.3367 & Sn, liq. \\
13.50 & 24000.00 & 7.890887 & 0.013547 & 255.8942 & 0.3439 & Sn, liq. \\
13.50 & 25000.00 & 8.310200 & 0.016070 & 259.8280 & 0.3939 & Sn, liq. \\
13.50 & 26000.00 & 8.751966 & 0.023433 & 263.9896 & 0.5522 & Sn, liq. \\
13.50 & 27000.00 & 9.208049 & 0.026346 & 268.4615 & 0.6288 & Sn, liq. \\
13.50 & 28000.00 & 9.650739 & 0.023145 & 272.4538 & 0.5555 & Sn, liq. \\
13.50 & 29000.00 & 10.067999 & 0.016308 & 275.7476 & 0.3744 & Sn, liq. \\
13.50 & 30000.00 & 10.548077 & 0.011604 & 280.1976 & 0.2792 & Sn, liq. \\
13.50 & 31000.00 & 11.009590 & 0.018451 & 284.1542 & 0.4378 & Sn, liq. \\
13.50 & 32000.00 & 11.478493 & 0.029895 & 288.0640 & 0.6972 & Sn, liq. \\
13.50 & 33000.00 & 11.935490 & 0.031608 & 291.5904 & 0.7116 & Sn, liq. \\
13.50 & 34000.00 & 12.426584 & 0.037159 & 295.5728 & 0.8106 & Sn, liq. \\
13.50 & 35000.00 & 12.964026 & 0.026347 & 300.5656 & 0.5936 & Sn, liq. \\
13.50 & 36000.00 & 13.410489 & 0.025393 & 303.7318 & 0.5550 & Sn, liq. \\
13.50 & 37000.00 & 13.956127 & 0.032009 & 308.5460 & 0.6923 & Sn, liq. \\
13.50 & 38000.00 & 14.411554 & 0.034617 & 311.4180 & 0.7603 & Sn, liq. \\
13.50 & 39000.00 & 14.942162 & 0.045336 & 315.4870 & 0.9486 & Sn, liq. \\
13.50 & 40000.00 & 15.437769 & 0.055283 & 318.7876 & 1.1464 & Sn, liq. \\
14.00 & 12000.00 & 4.004563 & 0.016634 & 234.7933 & 0.4267 & Sn, liq. \\
14.00 & 13000.00 & 4.315310 & 0.013553 & 238.2860 & 0.3455 & Sn, liq. \\
14.00 & 14000.00 & 4.695794 & 0.009232 & 243.4156 & 0.2558 & Sn, liq. \\
14.00 & 15000.00 & 5.038390 & 0.013430 & 247.4103 & 0.3528 & Sn, liq. \\
14.00 & 16000.00 & 5.407331 & 0.015364 & 251.8002 & 0.4016 & Sn, liq. \\
14.00 & 17000.00 & 5.768437 & 0.018984 & 255.8251 & 0.4818 & Sn, liq. \\
14.00 & 18000.00 & 6.078026 & 0.020213 & 258.4744 & 0.5064 & Sn, liq. \\
14.00 & 19000.00 & 6.526024 & 0.015287 & 264.3012 & 0.3753 & Sn, liq. \\
14.00 & 20000.00 & 6.902111 & 0.032592 & 268.3570 & 0.7962 & Sn, liq. \\
14.00 & 21000.00 & 7.295322 & 0.024778 & 272.2794 & 0.6038 & Sn, liq. \\
14.00 & 22000.00 & 7.722801 & 0.018860 & 277.1716 & 0.4876 & Sn, liq. \\
14.00 & 23000.00 & 8.072746 & 0.022101 & 279.8514 & 0.5296 & Sn, liq. \\
14.00 & 24000.00 & 8.528633 & 0.017606 & 284.9320 & 0.4117 & Sn, liq. \\
14.00 & 25000.00 & 8.922798 & 0.025345 & 288.2389 & 0.5900 & Sn, liq. \\
14.00 & 26000.00 & 9.341264 & 0.015144 & 292.3808 & 0.3643 & Sn, liq. \\
14.00 & 27000.00 & 9.790005 & 0.020808 & 296.7193 & 0.5121 & Sn, liq. \\
14.00 & 28000.00 & 10.212521 & 0.020145 & 300.2945 & 0.4806 & Sn, liq. \\
14.00 & 29000.00 & 10.678831 & 0.013619 & 304.8531 & 0.2950 & Sn, liq. \\
14.00 & 30000.00 & 11.133687 & 0.025237 & 308.9744 & 0.5331 & Sn, liq. \\
14.00 & 31000.00 & 11.559503 & 0.019524 & 311.8089 & 0.4861 & Sn, liq. \\
14.00 & 32000.00 & 12.041797 & 0.024809 & 316.3565 & 0.5625 & Sn, liq. \\
14.00 & 33000.00 & 12.481815 & 0.017466 & 319.6708 & 0.3794 & Sn, liq. \\
14.00 & 34000.00 & 13.042225 & 0.025816 & 325.6239 & 0.5487 & Sn, liq. \\
14.00 & 35000.00 & 13.490160 & 0.026697 & 328.8127 & 0.5942 & Sn, liq. \\
14.00 & 36000.00 & 13.976973 & 0.023784 & 332.5400 & 0.5074 & Sn, liq. \\
14.00 & 37000.00 & 14.470441 & 0.045255 & 336.5893 & 1.0513 & Sn, liq. \\
14.00 & 38000.00 & 15.013894 & 0.031595 & 341.2415 & 0.7168 & Sn, liq. \\
14.00 & 39000.00 & 15.461843 & 0.063489 & 344.0139 & 1.4318 & Sn, liq. \\
14.00 & 40000.00 & 16.045896 & 0.042029 & 349.8404 & 0.9842 & Sn, liq. \\
14.50 & 12000.00 & 4.582912 & 0.007015 & 262.1011 & 0.1886 & Sn, liq. \\
14.50 & 13000.00 & 4.937393 & 0.017456 & 266.8270 & 0.4647 & Sn, liq. \\
14.50 & 14000.00 & 5.292630 & 0.011201 & 271.4652 & 0.2945 & Sn, liq. \\
14.50 & 15000.00 & 5.645561 & 0.009642 & 275.8260 & 0.2559 & Sn, liq. \\
14.50 & 16000.00 & 6.016510 & 0.014799 & 280.4525 & 0.3866 & Sn, liq. \\
14.50 & 17000.00 & 6.316768 & 0.021216 & 283.0690 & 0.5591 & Sn, liq. \\
14.50 & 18000.00 & 6.730226 & 0.014224 & 288.4337 & 0.3641 & Sn, liq. \\
14.50 & 19000.00 & 7.132308 & 0.020820 & 293.2898 & 0.5243 & Sn, liq. \\
14.50 & 20000.00 & 7.536297 & 0.013867 & 297.9617 & 0.3677 & Sn, liq. \\
14.50 & 21000.00 & 7.933773 & 0.021921 & 302.2864 & 0.5628 & Sn, liq. \\
14.50 & 22000.00 & 8.335195 & 0.019873 & 306.5897 & 0.5011 & Sn, liq. \\
14.50 & 23000.00 & 8.683242 & 0.013042 & 309.3775 & 0.3075 & Sn, liq. \\
14.50 & 24000.00 & 9.102184 & 0.012355 & 313.5723 & 0.3428 & Sn, liq. \\
14.50 & 25000.00 & 9.565255 & 0.021404 & 319.0003 & 0.5468 & Sn, liq. \\
14.50 & 26000.00 & 9.964053 & 0.015277 & 322.4767 & 0.3925 & Sn, liq. \\
14.50 & 27000.00 & 10.408139 & 0.019740 & 326.8830 & 0.4950 & Sn, liq. \\
14.50 & 28000.00 & 10.840036 & 0.017365 & 330.8232 & 0.4325 & Sn, liq. \\
14.50 & 29000.00 & 11.218068 & 0.027258 & 333.4370 & 0.6788 & Sn, liq. \\
14.50 & 30000.00 & 11.720276 & 0.030036 & 338.5723 & 0.7006 & Sn, liq. \\
14.50 & 31000.00 & 12.165622 & 0.043037 & 342.4442 & 1.0157 & Sn, liq. \\
14.50 & 32000.00 & 12.667698 & 0.023781 & 347.5370 & 0.5507 & Sn, liq. \\
14.50 & 33000.00 & 13.082951 & 0.026961 & 350.2787 & 0.6444 & Sn, liq. \\
14.50 & 34000.00 & 13.559696 & 0.035368 & 354.6054 & 0.8685 & Sn, liq. \\
14.50 & 35000.00 & 14.096980 & 0.028031 & 359.5290 & 0.7128 & Sn, liq. \\
14.50 & 36000.00 & 14.589731 & 0.044928 & 363.9389 & 1.0866 & Sn, liq. \\
14.50 & 37000.00 & 15.048731 & 0.032162 & 367.2713 & 0.7033 & Sn, liq. \\
14.50 & 38000.00 & 15.550744 & 0.035261 & 371.2555 & 0.7974 & Sn, liq. \\
14.50 & 39000.00 & 16.084960 & 0.059644 & 375.9744 & 1.3244 & Sn, liq. \\
14.50 & 40000.00 & 16.654874 & 0.033684 & 381.2377 & 0.7290 & Sn, liq. \\
15.00 & 12000.00 & 5.239390 & 0.013665 & 292.6380 & 0.3456 & Sn, liq. \\
15.00 & 13000.00 & 5.556495 & 0.011242 & 296.5354 & 0.3298 & Sn, liq. \\
15.00 & 14000.00 & 5.930396 & 0.017716 & 301.7407 & 0.4907 & Sn, liq. \\
15.00 & 15000.00 & 6.284514 & 0.017718 & 306.2174 & 0.4901 & Sn, liq. \\
15.00 & 16000.00 & 6.602185 & 0.013503 & 309.5318 & 0.3505 & Sn, liq. \\
15.00 & 17000.00 & 7.029032 & 0.014551 & 315.5346 & 0.3659 & Sn, liq. \\
15.00 & 18000.00 & 7.357669 & 0.020777 & 318.9022 & 0.5440 & Sn, liq. \\
15.00 & 19000.00 & 7.719538 & 0.026584 & 322.8074 & 0.6786 & Sn, liq. \\
15.00 & 20000.00 & 8.196538 & 0.026469 & 329.5572 & 0.6880 & Sn, liq. \\
15.00 & 21000.00 & 8.578672 & 0.027176 & 333.7037 & 0.7112 & Sn, liq. \\
15.00 & 22000.00 & 8.933430 & 0.029838 & 336.8847 & 0.7391 & Sn, liq. \\
15.00 & 23000.00 & 9.310148 & 0.020309 & 340.2428 & 0.5301 & Sn, liq. \\
15.00 & 24000.00 & 9.753736 & 0.012579 & 345.4284 & 0.3228 & Sn, liq. \\
15.00 & 25000.00 & 10.163655 & 0.016098 & 349.5021 & 0.4071 & Sn, liq. \\
15.00 & 26000.00 & 10.561555 & 0.016104 & 353.2210 & 0.3813 & Sn, liq. \\
15.00 & 27000.00 & 11.014396 & 0.019411 & 358.0155 & 0.4858 & Sn, liq. \\
15.00 & 28000.00 & 11.416771 & 0.023859 & 361.2947 & 0.6106 & Sn, liq. \\
15.00 & 29000.00 & 11.935960 & 0.018195 & 367.3782 & 0.4574 & Sn, liq. \\
15.00 & 30000.00 & 12.363566 & 0.025636 & 371.0630 & 0.6492 & Sn, liq. \\
15.00 & 31000.00 & 12.866755 & 0.028633 & 376.3065 & 0.7265 & Sn, liq. \\
15.00 & 32000.00 & 13.375874 & 0.033752 & 381.5801 & 0.7584 & Sn, liq. \\
15.00 & 33000.00 & 13.796051 & 0.018316 & 384.6439 & 0.4619 & Sn, liq. \\
15.00 & 34000.00 & 14.271286 & 0.032851 & 388.7398 & 0.8052 & Sn, liq. \\
15.00 & 35000.00 & 14.765281 & 0.034918 & 393.2080 & 0.7936 & Sn, liq. \\
15.00 & 36000.00 & 15.254775 & 0.030343 & 397.6500 & 0.6816 & Sn, liq. \\
15.00 & 37000.00 & 15.759093 & 0.025198 & 401.7864 & 0.6278 & Sn, liq. \\
15.00 & 38000.00 & 16.203336 & 0.034807 & 404.8564 & 0.7483 & Sn, liq. \\
15.00 & 39000.00 & 16.709119 & 0.045350 & 409.0046 & 1.0015 & Sn, liq. \\
15.00 & 40000.00 & 17.300742 & 0.039786 & 414.9578 & 0.9597 & Sn, liq. \\
15.50 & 12000.00 & 5.890876 & 0.017640 & 324.3202 & 0.4860 & Sn, liq. \\
15.50 & 13000.00 & 6.260765 & 0.015980 & 329.7561 & 0.4508 & Sn, liq. \\
15.50 & 14000.00 & 6.601620 & 0.012750 & 334.1011 & 0.3351 & Sn, liq. \\
15.50 & 15000.00 & 6.944910 & 0.014567 & 338.4469 & 0.4078 & Sn, liq. \\
15.50 & 16000.00 & 7.301528 & 0.017838 & 342.8766 & 0.4839 & Sn, liq. \\
15.50 & 17000.00 & 7.674644 & 0.020470 & 347.7438 & 0.5463 & Sn, liq. \\
15.50 & 18000.00 & 8.068483 & 0.018058 & 352.7632 & 0.4734 & Sn, liq. \\
15.50 & 19000.00 & 8.435599 & 0.029300 & 356.8992 & 0.7607 & Sn, liq. \\
15.50 & 20000.00 & 8.818009 & 0.026390 & 361.2162 & 0.6961 & Sn, liq. \\
15.50 & 21000.00 & 9.246835 & 0.031620 & 366.7504 & 0.8375 & Sn, liq. \\
15.50 & 22000.00 & 9.574811 & 0.016744 & 369.4487 & 0.4245 & Sn, liq. \\
15.50 & 23000.00 & 9.987841 & 0.015967 & 373.9510 & 0.4373 & Sn, liq. \\
15.50 & 24000.00 & 10.404700 & 0.024327 & 378.5726 & 0.6227 & Sn, liq. \\
15.50 & 25000.00 & 10.878299 & 0.017398 & 384.3097 & 0.4459 & Sn, liq. \\
15.50 & 26000.00 & 11.273211 & 0.019309 & 387.8429 & 0.5054 & Sn, liq. \\
15.50 & 27000.00 & 11.671612 & 0.016268 & 391.3259 & 0.4385 & Sn, liq. \\
15.50 & 28000.00 & 12.172118 & 0.016902 & 397.3977 & 0.4386 & Sn, liq. \\
15.50 & 29000.00 & 12.578995 & 0.023057 & 400.6931 & 0.5580 & Sn, liq. \\
15.50 & 30000.00 & 13.036629 & 0.035681 & 405.4156 & 0.8883 & Sn, liq. \\
15.50 & 31000.00 & 13.473088 & 0.021304 & 409.0829 & 0.5305 & Sn, liq. \\
15.50 & 32000.00 & 13.929220 & 0.035625 & 413.3568 & 0.8695 & Sn, liq. \\
15.50 & 33000.00 & 14.502850 & 0.032632 & 420.0162 & 0.8089 & Sn, liq. \\
15.50 & 34000.00 & 14.881349 & 0.030835 & 422.1764 & 0.7365 & Sn, liq. \\
15.50 & 35000.00 & 15.452971 & 0.039165 & 428.6627 & 0.9305 & Sn, liq. \\
15.50 & 36000.00 & 15.919972 & 0.030601 & 432.0734 & 0.7162 & Sn, liq. \\
15.50 & 37000.00 & 16.398823 & 0.046140 & 435.9163 & 1.0870 & Sn, liq. \\
15.50 & 38000.00 & 16.936166 & 0.041490 & 441.3354 & 0.9635 & Sn, liq. \\
15.50 & 39000.00 & 17.417838 & 0.038525 & 445.2359 & 0.9009 & Sn, liq. \\
15.50 & 40000.00 & 17.856602 & 0.053280 & 447.3523 & 1.2051 & Sn, liq. \\
16.00 & 12000.00 & 6.588552 & 0.010326 & 358.5386 & 0.2844 & Sn, liq. \\
16.00 & 13000.00 & 6.963115 & 0.008863 & 364.2064 & 0.2645 & Sn, liq. \\
16.00 & 14000.00 & 7.296499 & 0.011164 & 368.5440 & 0.3202 & Sn, liq. \\
16.00 & 15000.00 & 7.667942 & 0.012257 & 373.7580 & 0.3362 & Sn, liq. \\
16.00 & 16000.00 & 7.986862 & 0.023726 & 377.2274 & 0.6304 & Sn, liq. \\
16.00 & 17000.00 & 8.387534 & 0.011216 & 382.9230 & 0.2994 & Sn, liq. \\
16.00 & 18000.00 & 8.784691 & 0.018839 & 388.1555 & 0.5334 & Sn, liq. \\
16.00 & 19000.00 & 9.127327 & 0.021727 & 391.6908 & 0.5879 & Sn, liq. \\
16.00 & 20000.00 & 9.533076 & 0.020942 & 396.7413 & 0.5673 & Sn, liq. \\
16.00 & 21000.00 & 9.933283 & 0.024231 & 401.6466 & 0.6539 & Sn, liq. \\
16.00 & 22000.00 & 10.334831 & 0.024931 & 406.1694 & 0.5965 & Sn, liq. \\
16.00 & 23000.00 & 10.720260 & 0.018324 & 410.1779 & 0.4803 & Sn, liq. \\
16.00 & 24000.00 & 11.176589 & 0.014799 & 415.8585 & 0.3876 & Sn, liq. \\
16.00 & 25000.00 & 11.556170 & 0.016564 & 419.4221 & 0.4253 & Sn, liq. \\
16.00 & 26000.00 & 11.969329 & 0.019789 & 423.5549 & 0.5300 & Sn, liq. \\
16.00 & 27000.00 & 12.386591 & 0.016447 & 427.6021 & 0.4415 & Sn, liq. \\
16.00 & 28000.00 & 12.822914 & 0.020080 & 431.9745 & 0.5029 & Sn, liq. \\
16.00 & 29000.00 & 13.382687 & 0.021465 & 439.2861 & 0.5656 & Sn, liq. \\
16.00 & 30000.00 & 13.718886 & 0.020638 & 441.0250 & 0.5136 & Sn, liq. \\
16.00 & 31000.00 & 14.243119 & 0.018333 & 447.2015 & 0.4761 & Sn, liq. \\
16.00 & 32000.00 & 14.708282 & 0.037634 & 451.6111 & 0.9578 & Sn, liq. \\
16.00 & 33000.00 & 15.212158 & 0.030973 & 456.7367 & 0.7536 & Sn, liq. \\
16.00 & 34000.00 & 15.622535 & 0.032029 & 459.6234 & 0.7620 & Sn, liq. \\
16.00 & 35000.00 & 16.115131 & 0.031839 & 464.3711 & 0.8149 & Sn, liq. \\
16.00 & 36000.00 & 16.555052 & 0.020730 & 467.7551 & 0.4639 & Sn, liq. \\
16.00 & 37000.00 & 17.138811 & 0.032858 & 474.1377 & 0.8112 & Sn, liq. \\
16.00 & 38000.00 & 17.546811 & 0.037717 & 476.1300 & 0.9359 & Sn, liq. \\
16.00 & 39000.00 & 18.094883 & 0.029588 & 481.5681 & 0.7611 & Sn, liq. \\
16.00 & 40000.00 & 18.686321 & 0.048779 & 488.1845 & 1.0903 & Sn, liq. \\
12.50 & 5000.00 & -4.980080 & 0.002119 & 247.1710 & 0.0645 & Fe, sol. \\
13.50 & 5000.00 & -4.181941 & 0.001539 & 337.7488 & 0.0441 & Fe, sol. \\
13.00 & 5000.00 & -1.083734 & 0.002473 & 186.3594 & 0.1085 & Cu, sol. \\
13.00 & 6000.00 & -0.760377 & 0.003942 & 195.4087 & 0.1721 & Cu, sol. \\
13.50 & 5000.00 & -0.793778 & 0.001635 & 216.3550 & 0.0776 & Cu, sol. \\
13.50 & 6000.00 & -0.477312 & 0.003917 & 225.4728 & 0.1695 & Cu, sol. \\
14.00 & 5000.00 & -0.467014 & 0.001887 & 249.6611 & 0.0901 & Cu, sol. \\
14.00 & 6000.00 & -0.143542 & 0.005534 & 259.3500 & 0.2615 & Cu, sol. \\
14.50 & 5000.00 & -0.108334 & 0.002295 & 286.0529 & 0.0963 & Cu, sol. \\
14.50 & 6000.00 & 0.204252 & 0.002030 & 295.5932 & 0.0933 & Cu, sol. \\
14.50 & 7000.00 & 0.532781 & 0.005487 & 305.4198 & 0.2508 & Cu, sol. \\
15.00 & 5000.00 & 0.280577 & 0.000841 & 325.6906 & 0.0538 & Cu, sol. \\
15.00 & 6000.00 & 0.590610 & 0.001362 & 335.3652 & 0.0681 & Cu, sol. \\
15.00 & 7000.00 & 0.917560 & 0.003443 & 345.4495 & 0.1676 & Cu, sol. \\
12.00 & 5000.00 & -5.302303 & 0.003458 & 208.0766 & 0.1526 & Fe, sol* \\
13.00 & 5000.00 & -4.607604 & 0.002558 & 290.3764 & 0.0785 & Fe, sol* \\
13.00 & 6000.00 & -4.162117 & 0.021780 & 303.7985 & 0.8596 & Fe, sol* \\
13.50 & 6000.00 & -3.782597 & 0.002581 & 349.4452 & 0.0848 & Fe, sol* \\
13.50 & 7000.00 & -0.088877 & 0.024718 & 237.4735 & 1.1242 & Cu, sol* \\
\end{longtable*}

\end{document}